\documentclass[aps,twocolumn,showpacs]{revtex4}

\usepackage{mathtools}
\usepackage{bbm}
\usepackage{longtable}
\usepackage{tcolorbox}
\usepackage{graphicx}
\usepackage{verbatim}
\usepackage{amssymb,amsfonts,amsmath}
\usepackage{fancyref}
\usepackage[utf8]{inputenc}
\usepackage{mathpazo}
\usepackage{xcolor}
\usepackage{hyperref}
\hypersetup{ 
	colorlinks   = true
}
\usepackage{algorithm}
\usepackage[noend]{algpseudocode}

\graphicspath{{FIGURES/}}

\definecolor{pastel_green}{rgb}{0.18,0.65,0.34}

\def \mbf{\mathbf}

\def \M{\mathcal{M}}  
\def \P{\mathbb{P}}
\def \E{\mathbb{E}}

\def\thet{{\theta}^{(t)}}
\def\thetp{{\theta}^{(t+1)}}
\def\thets{{\theta}^{(t)\star}}

\def\be{\begin{equation}} 
\def\ee{\end{equation}} 
\newcommand \bea {\begin{eqnarray}} 
\newcommand \eea {\end{eqnarray}} 

\newcommand{\nn} {\nonumber}

\def\bd{\boldsymbol}

\begin{document}

\title{The strength of protein-protein interactions controls the information capacity and dynamical response of signaling networks}

\author{Ching-Hao Wang}
\email{chinghao@bu.edu}
\affiliation{Department of Physics and Biological Design Center, Boston University, Boston, MA 02215, USA}

\author{Caleb J. Bashor}
\email{caleb.bashor@rice.edu}
\affiliation{Department of Bioengineering, Rice University, Huston, TX 77030, USA}

\author{Pankaj Mehta}
\email{pankajm@bu.edu}
\affiliation{Department of Physics and Biological Design Center, Boston University, Boston, MA 02215, USA}

\date{\today}


\begin{abstract}
	Eukaryotic cells transmit information by signaling through complex networks of interacting proteins. Here we develop a theoretical and computational framework that relates the biophysics of protein-protein interactions (PPIs) within a signaling network to its information processing properties. To do so, we generalize statistical physics-inspired models for protein binding to account for interactions that depend on post-translational state (e.g. phosphorylation). By combining these models with information-theoretic methods, we find that PPIs are a key determinant of information transmission within a signaling network, with weak interactions giving rise to "noise" that diminishes information transmission. While noise can be mitigated by increasing interaction strength, the accompanying increase in transmission comes at the expense of a slower dynamical response. This suggests that the biophysics of signaling protein interactions give rise to a fundamental ``speed-information" trade-off. Surprisingly, we find that cross-talk between pathways in complex signaling networks do not significantly alter information capacity--an observation that may partially explain the promiscuity and ubiquity of weak PPIs in heavily interconnected networks. We conclude by showing how our framework can be used to design synthetic biochemical networks that maximize information transmission, a  procedure we dub "InfoMax" design. 
\end{abstract}

\maketitle

\section*{Introduction}

Cells have evolved complex protein signaling networks to process information about their living environments \cite{barabasi2004network, blais2005constructing, macarthur2009systems, martello2014nature}. These networks play a central role in cellular decision-making, development, growth, and migration \cite{seet2006reading, scott2009cell, lim2014cell}. In eukaryotic cells, signaling pathways such as Wnt/$\beta$-Catenin\cite{angers2009proximal,macdonald2009wnt} and TGF-$\beta$ pathways\cite{massague2012tgfbeta} have important homeostatic functions (e.g., cell proliferation, differentiation, and fate determination), with disruptions in their signaling leading to tumorigenesis and drive metastasis \cite{anastas2013wnt,moustakas2014tgfbeta}.

Information transfer in signaling networks occurs via the addition of covalent chemical groups that alter the regulatory state of a signaling protein (e.g. phosphorlyation of a Tyrosine residue). Addition and removal these post-translational modifications (PTMs) are respectively catalyzed by "writer" (e.g. a kinase) and "eraser" (e.g. a phosphatase) enzyme activities. Information transfer occurs when the ratio of these opposing activities is altered by an upstream input (e.g. ligand binding to receptor), and becomes rapidly and reversibly encoded in the PTM state of the downstream substrate (e.g. phosphorylated or non-phosphorylated). An important breakthrough in the understanding of signaling network connectivity came with the discovery of protein-protein interaction (PPI) domains that specifically bind to PTM-modified motifs, effectively “decoding” the PTM state of a substrate\cite{deribe2010post, scott2009cell}. By linking an activity to a substrate through binding, PTM-mediated PPI interactions serve as signaling network links by interconnecting writer/eraser cycles (Fig. ~\ref{fig:network-model}A). One of the best-known examples of a PTM-binding domain is the Src homology 2 (SH2) domain, which specifically docks to motifs containing phosphorylated tyrosine. For example, SH2 recognition plays a central role in the EGF pathway, connecting initial receptor autophosphorylation to downstream signaling events via recruitment of SH2 domain-containing enzymes to their substrates (Fig.~\ref{fig:network-model}A,C).

Given their role in mediating information transfer between signaling proteins, the question naturally arises as to how the biophysical features of PTM-PPI interactions relate to a pathway’s emergent, network-level information processing properties. Here, we create a theoretical framework for exploring this relationship using a thermodynamically-inspired statistical model in which biochemical partition functions relate the probability of finding the system in a given state (e.g. bound, unbound, etc.) to relevant biophysical features like interaction affinity and species concentration \cite{ackers1982quantitative,hill2013cooperativity, weinert2014scaling}. Models of this class have been successfully used to understand the biophysics of promoter regulation in transcriptional networks\cite{bintu2005transcriptional, kinney2010using, garcia2010transcription, weinert2014scaling}. Here, we extend this approach to signaling networks by introducing variables representing PTM-dependent PPIs, thereby accounting for the non-equilibrium nature of reversible, enzyme-catalyzed phosphorylation. We combine this statistical physics approach with information theory\cite{shannon2001mathematical,cover2012elements}, which has seen widespread recent application in biology\cite{johnson1970information}. Examples include the modeling of information processing in gene networks\cite{tkavcik2009optimizing, walczak2010optimizing,tkavcik2011information, granados2018distributed}, enzyme cascades \cite{detwiler2000engineering}, and bacterial signaling networks\cite{mehta2009information, tostevin2009mutual}, as well as calculating information capacity in canonical eukaryotic signaling networks from single cell measurement of input-output relationships\cite{cheong2011information, brennan2012information}.

Our joint framework allows us to investigate the relationship between the biophysics of PPI-PTM interactions and signaling network information processing. We chose to model a simple, idealized signaling pathway in order to more directly probe this relationship. Here, our approach is inspired by synthetic biology, where a principle goal is engineering synthetic regulatory circuits capable of executing designed regulatory function, typically through direct experimental manipulating features like protein expression level and PPI strength. Thus, in contrast to previous approaches that investigate the information capacity of pre-existing, native networks, our goal with this work is to ask how we can manipulate the biophysics of PPIs to engineer new networks that optimize information transmission. Information processing circuits must necessarily balance three competing requirements that are often in tension: i) minimizing unwanted “noise” that corrupts the true signal, ii) ensuring that the circuits can respond quickly to dynamical perturbations, and iii) maximizing the dynamic range of inputs. In signaling networks, it has been argued significant noise is introduced by weak, promiscuous PPIs\cite{ladbury2012noise, voliotis2014information}, often in combination with low levels of background kinase and phosphatase activity\cite{chung2010spatial,schlessinger2000cell}.Thus, we hypothesize in the current work that while increasing the strength of PTM-PPI interactions may reduce noise, it may also involve inherent tradeoffs in response times and dynamic range.

Motivated by these considerations, we focus in this article on a series of interrelated conceptual questions: How can we quantify noise due to promiscuous PPIs? How does the strength of PPIs affect information transmission and dynamic response times in signaling networks? How do network architecture and cross-talk affect information transmission\cite{hill1998receptor,schwartz2002networks, hunter2007age, voliotis2014information, kontogeorgaki2017noise}? Can we rationally choose PPIs in synthetic biochemical networks that maximize information transmission? We begin by discussing how to generalize thermodynamic models to binding that include PTMs. We then discuss how basic elements of these models can serve as an input into information theoretic calculations. Using this framework, we quantitatively show how weak PPIs give rise to non-specific binding, resulting in ``noise'' that reduces information transmission. We then show that while noise can be diminished by increasing PPI strength, increased information transmission comes results in a slower dynamical response—a biophysical manifestation of what in engineering is often called the “gain-bandwidth” tradeoff. We then show that cross-talk between pathways in highly interconnected signaling networks does not significantly alter information capacity. We conclude by discussing "InfoMax", a new procedure for designing synthetic biochemical networks that optimize gain-bandwidth tradeoff.

\section*{Including post-translational modifications in thermodynamic models}

To construct a thermodynamic model, we consider an idealized post-translational signaling network with phosphorylation as the only PTM. Each node in the network represents a distinct kinase activity, and linkages between nodes are mediated by PTM-dependent PPIs (Figure~\ref{fig:network-model}A). Here, phosphorylation of a kinase node by an upstream activity renders it `active' and competent to engage with and phosphorylate (and subsequently activate) a downstream kinase. We sought to create a generalizable thermodynamic expression for describing such a network.

For a given multi-state molecular system, thermodynamics provide a concise description of the statistical weight of each state, and therefore the probability of observing a state when the system is at steady state. At thermal equilibrium the statistical weight of a given microscopic configuration is proportional to its Boltzmann factor defined as $e^{-\beta E}$, where $E$ is the energy of this microstate and $\beta = 1/(k_BT)$ is the inverse temperature with $k_B$ being the Boltzmann constant. As we noted, conventional thermodynamic prescription based on transcriptional regulation\cite{bintu2005transcriptional, kinney2010using, garcia2010transcription, weinert2014scaling} does not include PTMs and PTM-dependent bindings. Here we introduce a new set of variables to account for PTMs.

For brevity, we consider a simplified signaling network (Figure~\ref{fig:network-model}C); a linear pathway consisting of a membrane-spanning receptor kinase  $R$,  a single freely-diffusing protein kinases $K_1$, and target transcription factor $TF$. We treat $R, K_1, TF\in\{0,1\}$ with value 1 indicating a phoshphorylated state (transcribed state for $TF$) and 0 otherwise. Pathway activation (input) is initiated by ligand ($L$) binding to the receptor at the cell surface, leading to receptor autophosphorylation (i.e. $R=1$). This results in phosphorylation-dependent recruitment and phosphorylation of $K_1$ (i.e. $K_1=1$). Phosphorylated $K_1$ then translocates into the nucleus where it binds to and phosphorylates $TF$ (i.e. $TF=1$), activating transcription. 

Within the context of this simplified signaling system, we begin to describe the thermodynamics of the interactions involved, breaking down the network depicted in Figure~\ref{fig:network-model}C into three parts and enumerating the possible states within each. As depicted in Figure~\ref{fig:network-model}A,C, the receptor kinase only has two possible PTM states (phosphorylated or not). We label the probability of phosphorylated receptor kinase as $\P(R=1)=q$, where $q\in [0,1]$ is the parameter that encapsulates ligand activation. The probability of the complementary configuration is therefore given by $\P(R=0)=1-q$. (ii) Based on our discussion above, the interaction between $R$ and $K$ depends crucially on the value of $R$. Simple enumeration reveals that there are four possible scenarios: $(K_1, R)\in\{(0, 0), (1, 0), (0, 1), (1, 1)\}$, as shown in Figure~\ref{fig:network-model}D. The first two involve the interaction between unphosphorylated receptor (i.e. $R=0$) and $K_1$ while the last two involve that between phosphorylated receptor (i.e. $R=1$) and $K_1$. Thermodynamics dictates that when a system reaches equilibrium, the steady-state distribution of a microscopic state is given by the Boltzmann factor of that state divided by the sum of the Boltzmann factor of all possible states (i.e., partition function). It's worth noting that although enzymatic reactions (i.e. phosphorylation) are involved in signaling can drive a system out of equilibrium, we show in Appendix that the steady state distribution of a given state takes the Boltzmann form. In other words, the probability of having a phosphorylated PK given that the receptor kinase is phosphorylated, \emph{viz.}  
\be\label{eq:PIOdef}
\P(K_1=1|R=1)=\frac{ e^{-\beta \theta_{R,K_1}}}{1+  e^{-\beta \theta_{R,K_1}}},
\ee
where the numerator is the Boltzmann factor associated with this configuration while the denominator is the sum of this factor and that associated with $(K_1, R)= (0,1)$ (i.e., factor 1). Here we denote $\theta_{R,K_1}$ as the binding affinity (BA) of $R$ to $K_1$ (i.e. $\theta_{R, K_1}=\Delta F - \tilde{\mu}$, where $\Delta F$ is the free energy difference between the bound and unbound state and $\tilde{\mu}$ is the chemical potential of phosphorylated $R$; see SI Section for its expression in terms of kinetic parameters.) By conservation, the complementary configuration has probability $\P(K_1=0|R=1)= 1- \P(K_1=1|R=1) = 1 / (1+ \exp(-\beta \theta_{R,K_1}))$. The cases where receptor is not phosphorylated (i.e. $R=0$) are similar except that the binding affinity is parameterized by $W$. Since $W$ controls the amount of low-probability, non-specific binding, we assume  $W$ is  positive and large (i.e. $\beta W\gtrsim 1$ ) so that the probability of having a phosphorylated $K_1$ given that there's no signal input is almost zero, \emph{viz}.
\be
\P(K_1=1|R=0) = \frac{e^{-\beta W}}{1+ e^{-\beta W}} \approx 0,
\ee
which implies $\P(K_1=0|R=0) = 1-\P(K_1=1|R=0) \approx 1$. Note that practically this would require $\beta W \ge 4.60$ in order to achieve $e^{-\beta W} \le 0.01$. With all these defined, one can summarize all four configurations and their statistical weights by the phosphorylation probability of $K_1$ conditioned on the state of $R$, $P(K_1|R)$ (see Fig.~\ref{fig:network-model}D). (iii) Finally, since the thermodynamic description of the interaction between $K_1$ and $TF$ is the same as that between $R$ and $K_1$, one can write down $\P(TF|K_1)$ in a similar fashion by relating $\theta_{R, K_1}$ to $\theta_{K_1, TF}$ (see Fig.~\ref{fig:network-model}D.)\\

\section*{Mutual information and PPIs}

Mutual information between two random variables measures how much knowing one tells us about the other, usually measured in units of bits\cite{shannon2001mathematical,cover2012elements}. In biology, it has been widely used to characterize the information transfer by biochemical systems \cite{johnson1970information, detwiler2000engineering, tkavcik2009optimizing, mehta2009information,walczak2010optimizing,tkavcik2011information,cheong2011information, brennan2012information}. Here we focus on defining this information-theoretic quantity in terms of PPIs for a given PK signaling network.


\begin{figure*}
	\centering
	\includegraphics[width=0.7\linewidth]{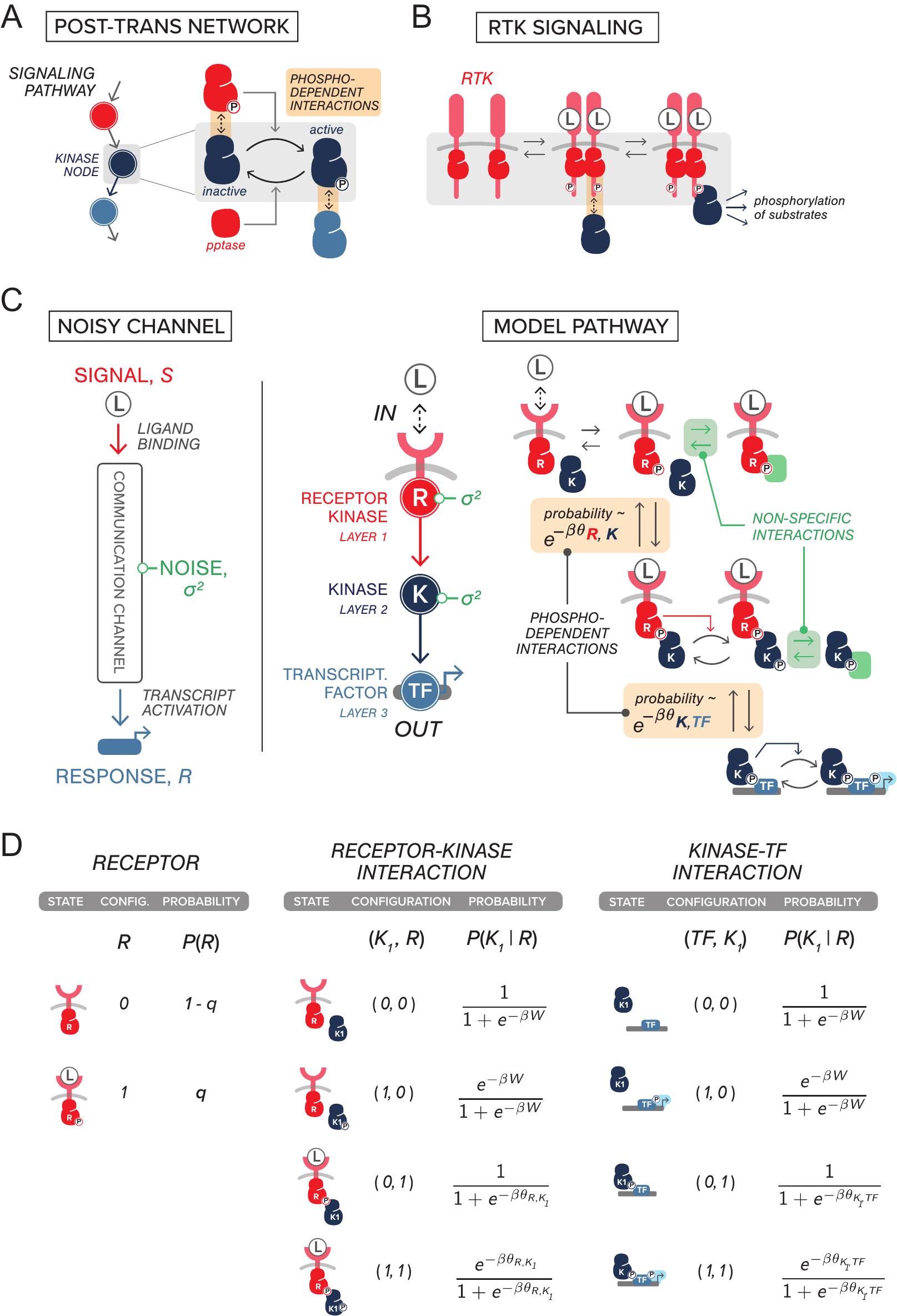}
	\caption{Relating the biophysics of post-translational modification (PTMs) and protein-protein interactions (PPIs) to network-level information properties ({\it A}) A model signaling network that consists of three nodes is shown. The PPIs and PTMs that involve a single node (e.g. kinase node, colored dark blue) is illustrated on the right. In a PTM network, edges between nodes encodes both the phosphorylation dependent PPIs and the resulting change in enzymatic activities (e.g. active/inactive).  ({\it B}) Naturally occurring pathways that can be conceptualized as the network shown in {\it A}. In receptor tyrosine kinase (RTK) signaling, binding of ligand (e.g. epidermal growth factor, EGF) to the extracellular domain of EGF receptor (EGFR, pink lobe) leads to the its dimerization and the phoshorylation of its kinase domain. This triggers signaling through phosphoryaltion-dependent interactions. ({\it C}) A signaling pathway can be viewed as a noisy communication channel (left). The input to this pathway is a ligand ($L$) that binds to the receptor kinase ($R$) which, through allosteric interactions, leads to receptor kinase phosphorylation. The phosphorylated receptor kinase then specifically binds its cognate kinase ($K$) which in term translocates into the nucleus to activate transcription. A pictorial summary of these events are shown on the right. The steady-state phosphorylation probability is annotated. Non-specific interactions (i.e. those highlighted in green) serve as noise in the network representation. All species are colored according to the nodes they correspond to (left). ({\it D}) Probability of PTM states in the thermodynamic model. Species are labeled with reference to {\it A}. As in the main text, binary variables $R, K_1, TF\in\{0,1\}$ are used to indicate the PTM states of these species with value 1 indicating a phosphorylated state (transcribed state for TF) and 0 otherwise. Panels are organized according to the binding interactions involved and are indicted at the top. In ${\it C, D}$, $\beta = 1/(k_BT)$ is the inverse temperature and $\theta_{i,j}$ is the binding affinity of $i$ to $j$ (see main text for definition).}
	\label{fig:network-model}
\end{figure*}

The mutual information of interest is that between the receptor kinase and TF output, $I(R;TF)$, since it quantifies how many input states cell can distinguish solely by examining its TF readout. Mathematically, 
\be\label{eq:MIdef}
I(R; TF) = \sum_{R}\sum_{TF} \P(R)\P(TF|R)\log_2\left[\frac{\P(TF|R)}{\P(TF)}\right].
\ee
Note that since the summations in Eq.\eqref{eq:MIdef} are over $\{0,1\}$, this signaling network represents a discrete (binary) channel\cite{cover2012elements}. Physically speaking, $\P(TF)$ quantifies the transcriptional readout, $\P(TF|R)$ defines the input-output relation (i.e., channel transfer function), and $\P(R)$ measures the input, all at steady-states. Note that the state of PK, $K_1$, is absent from this expression since it is embedded in the input-output relation. Within the thermodynamic framework defined based on Fig.~\ref{fig:network-model}C and detailed in Fig.~\ref{fig:network-model}D, all quantities in Eq.\eqref{eq:MIdef} can be explicitly calculated: signal input $\P(R)$ is given in Fig.~\ref{fig:network-model}D while the channel input-output relation (i.e. transfer function), $\P(TF|R)$, is obtained by first invoking the conditional independence of $TF$ and $R$ on $K_1$, then marginalizing contributions from $K_1$, \emph{viz}. $\P(TF|R)=\sum_{K_1}\P(TF|K_1)\P(K_1|R)$. Finally, the output is simply given by $\P(TF)=\sum_R \P(TF|R)\P(R)$. Explicitly, the transfer function is given by:
\begin{widetext}
\bea \label{eq:Q1}
\P(TF=1| R=1) &=& \P(TF=1|K_1=1)\P(K_1=1|R=1) + \P(TF=1|K_1=0)\P(K_1=0|R=1)\nonumber\\
&=& \left(\frac{e^{-\beta \theta_{K_1, TF}}}{1+ e^{-\beta \theta_{K_1, TF}}}\right) \left(\frac{e^{-\beta \theta_{R, K_1}}}{1+ e^{-\beta \theta_{R, K_1}}}\right) +\left(\frac{e^{-\beta W}}{1+ e^{-\beta W}}\right) \left(\frac{1}{1+ e^{-\beta \theta_{R, K_1}}}\right)\\
&\approx& \left(\frac{e^{-\beta \theta_{K_1, TF}}}{1+ e^{-\beta \theta_{K_1, TF}}}\right) \left(\frac{e^{-\beta \theta_{R, K_1}}}{1+ e^{-\beta \theta_{R, K_1}}}\right)
\eea
\bea
\P(TF=1| R=0) &=& \P(TF=1|K_1=1)\P(K_1=1|R=0) + \P(TF=1|K_1=0)\P(K_1=0|R=0)\nonumber\\
&=& \left(\frac{e^{-\beta \theta_{K_1, TF}}}{1+ e^{-\beta \theta_{K_1, TF}}}\right) \left(\frac{e^{-\beta W}}{1+ e^{-\beta W}}\right) +\left(\frac{e^{-\beta W}}{1+ e^{-\beta W}}\right) \left(\frac{1}{1+ e^{-\beta W}}\right)\\
&\approx& 0
\eea
\bea
\P(TF=0| R=1) &=& \P(TF=0|K_1=1)\P(K_1=1|R=1) + \P(TF=0|K_1=0)\P(K_1=0|R=1)\nonumber\\
&=& \left(\frac{1}{1+ e^{-\beta \theta_{K_1, TF}}}\right) \left(\frac{e^{-\beta \theta_{R, K_1}}}{1+ e^{-\beta \theta_{R, K_1}}}\right) +\left(\frac{1}{1+ e^{-\beta W}}\right) \left(\frac{1}{1+ e^{-\beta \theta_{R, K_1}}}\right)\\
&\approx& \left(\frac{1}{1+ e^{-\beta \theta_{K_1, TF}}}\right) \left(\frac{e^{-\beta \theta_{R, K_1}}}{1+ e^{-\beta \theta_{R, K_1}}}\right) + \left(\frac{1}{1+ e^{-\beta \theta_{R, K_1}}}\right)
\eea
\bea
\P(TF=0| R=0) &=& \P(TF=0|K_1=1)\P(K_1=1|R=0) + \P(TF=0|K_1=0)\P(K_1=0|R=0)\nonumber\\
&=& \left(\frac{1}{1+ e^{-\beta \theta_{K_1, TF}}}\right) \left(\frac{e^{-\beta W}}{1+ e^{-\beta W}}\right) +\left(\frac{1}{1+ e^{-\beta W}}\right) \left(\frac{1}{1+ e^{-\beta W}}\right)\\
&\approx& 1,
\eea
\end{widetext}
where the approximation in the last line of these expressions indicates the limit where $\beta W\gtrsim 1$ so that $e^{-\beta W} \rightarrow 0$. In this limit, the output is simply
\begin{widetext}
\bea
\P(TF=1) &\approx& \left(\frac{e^{-\beta \theta_{K_1, TF}}}{1+ e^{-\beta \theta_{K_1, TF}}}\right) \left(\frac{e^{-\beta \theta_{R, K_1}}}{1+ e^{-\beta \theta_{R, K_1}}}\right) q\\
\P(TF=0) &\approx& \left[\left(\frac{1}{1+ e^{-\beta \theta_{K_1, TF}}}\right) \left(\frac{e^{-\beta \theta_{R, K_1}}}{1+ e^{-\beta \theta_{R, K_1}}}\right) + \left(\frac{1}{1+ e^{-\beta \theta_{R, K_1}}}\right)\right] q +  (1-q)
\eea
\end{widetext}
With all these at hand, we can express Eq. \eqref{eq:MIdef} as a function of BAs $\theta_{i,j}$. In SI Section 2, we provide the analytic expression of mutual information Eq.\eqref{eq:MIdef} in terms of BAs. We have thus established an explicit functional relation between mutual information and PPIs.

\section*{Results}
\subsection*{Weak binding affinities result in noise that limit the signal-to-noise ratio and information capacity}

\begin{figure*}
	\centering
	\includegraphics[width=\linewidth]{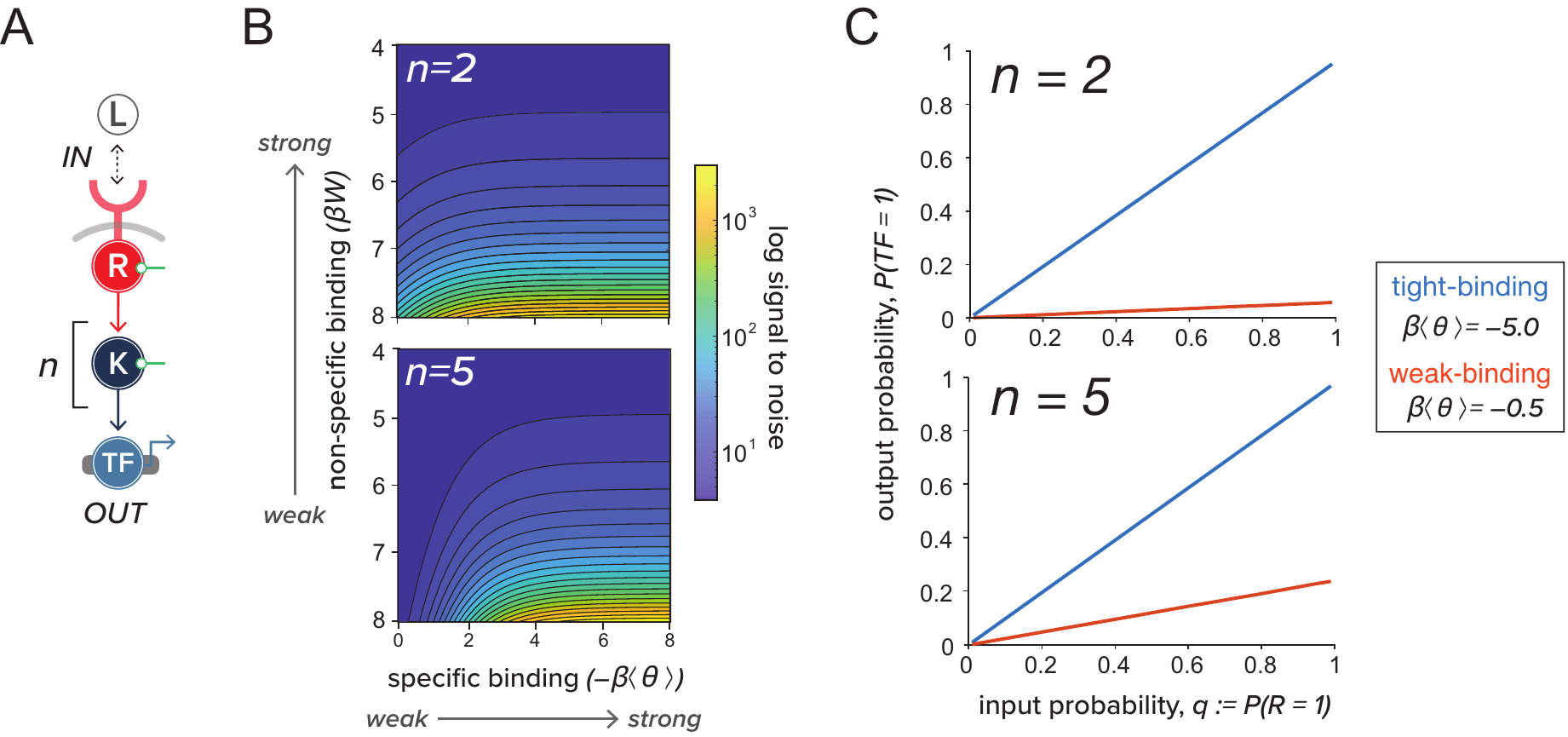}
	\caption{Noise due to non-specific protein-protein interactions (PPIs) limits the quality of information transmission.  ({\it A}) A simple linear network that mediates information of input (L) through a $n$-layer kinase cascade ( $K_i,\, i = 1, 2,\cdots, n$), to an output transcription factor (TF) which is active when phosphorylated. As in Figure~\ref{fig:network-model} {\it C}, green circles indicate noise. ({\it B})  Color map shows the numerically simulated log-signal-to-noise ratio (log-SNR), defined in Eq.\eqref{eq:SNRtb}, of the network shown in {\it A} at different level of specific and non-specific interactions. Binding affinities $\beta\theta_{i,j}$ is drawn from a normal distribution with mean $\langle\theta\rangle$ and variance $\sigma^2=0.01$ (see main text for simulation details). This quantity can also be obtained by solving Eq.\eqref{eq:SI:CVnum}\eqref{eq:SI:CVden} (see SI Section~\ref{sec:SI-SNR}). In this panel, we show the result for $n=2,5$. ({\it C}) Input-output relation of the $n$-layer kinase cascade ($n=2,5$) for tight- and weak-binding with $\beta W =2.0$.}
	\label{fig:SNR-all}
\end{figure*}

A  key biophysical quantity that controls the network level properties is the binding affinity -- or equivalently the binding energy -- between proteins. When the binding affinity is large, proteins stay tightly bound to there targets. Small binding affinities allow proteins to quickly bind and unbind from targets but can give rise to transient binding. Here we examine how these considerations affect information transmission through a signaling network. To understand this tradeoff quantitatively, we consider a family of single-input, single-output signaling networks consisting of a receptor kinase $R$ that phosphorylates a variable size intermediate layer consisting of $n$ kinases $K_i$ $(i=1,\cdots, n)$, and a transcriptional output TF (see Fig.~\ref{fig:SNR-all}A).  The binary variables $R, K_i, TF\in\{0,1\}$ encode the PTM-state of the protein with the value 1 indicating a phosphorylated state and 0 an unphosphorlyated state. We assume that the output transcription factor is active if and only if it is phosphorylated and the that the circuit is designed to activate the TF in the presence of a ligand at concentration $L$. We focus on information transmission at steady-state and neglect information encoded in the temporal dynamics. 

A fundamental measure of noise in signaling networks is the signal-to-noise ratio (SNR) \cite{detwiler2000engineering, cover2012elements}. To define the SNR, we make use of the probability that the output TF is active in the presence of the ligand $Q(L)\equiv P(TF=1|L)$. In general, this input-output function is probabilistic. The stochasticity in $Q$ stems from the probabilistic nature of protein-protein binding that is inherent in our thermodynamically-inspired models. And as in all thermodynamic models the more negative the binding affinities ($\theta_{k,j}$ where $k,j \in \{R, K_i, TF\}$), the smaller the effect of thermal fluctuations. In terms of $Q(L)$, the output obtained under a high input, $L=1$, (e.g. large number of phosphorylated receptor kinase) defines the best ``signal'' one can obtain for a given realization of BAs. On the other hand, there can still output signals even when the input is absent (i.e. $L=0$) due to thermal ``noise'' inherent in PPIs (i.e. contributions from $W$, see SI Section 1 for details).  We therefore define the signal-to-noise ratio (SNR) of a given network/channel as the ratio between $Q(L=1)$ and $Q(L=0)$, averaged over realizations of BAs.

To understand the effect of the strength of PPI on the SNR, we consider drawing the binding affinities for the interactions in our network from a normal distribution $\theta_{i,j}\sim \mathcal{N}(\mu,\sigma)$ with mean binding affinity $\mu\equiv \langle \theta\rangle$ and variance $\sigma^2$, where $\langle\cdot\rangle$ refers to average over different realizations of BAs. The PPIs involving $W$, which sets the time scale of unbinding between unphosphorylated kinase to its substrate, is varied in the following analysis.  This allows us to probe the effect of both the mean binding strength as well as the thermal noise resulting from $W$.  Under these assumptions, we can analytically derive a formula for the SNR (see SI Section 1 for full derivations). When proteins bind tightly (i.e. large negative binding energies $ \beta\mu \lesssim -1$),  the SNR for the simplest signaling network $ L\rightarrow R\rightarrow TF$ reduces to the following simple expression:
\be\label{eq:SNRtb}
\text{SNR} \equiv  \frac{\langle Q(L=1)\rangle}{\langle Q(L=0) \rangle } =e^{\beta W}\left[1-e^{\beta \left(\mu+\frac{\sigma^2}{2}\right)}\right]
\ee

For networks with $n$-layers of kinase between input $R$ and output $TF$, as depicted in Figure~\ref{fig:SNR-all}A, we plot the color map of their log-SNR at different level of specific and non-specific PPIs in Figure~\ref{fig:SNR-all}B. Regardless of the depth of network, $n$, strong specificity in PPIs, namely, tighter binding, always leads to higher SNR. This suggests that BA is an important source of ``noise'' that limits the resolution of output signal. To further explore this idea, we calculate the corresponding input-output relation (i.e. $\P(TF=1)$ as a function of $\P(R=1)\equiv q$) in Figure~\ref{fig:SNR-all}C both at tight- and weak-binding. As shown, networks with strong BAs always have a larger gain, implying a higher information capacity\cite{cover2012elements,detwiler2000engineering}.
Note that the activation of the receptor kinase, $R=1$, depends on whether it is bound to ligand, and thus $q$ is implicitly a function of ligand concentration $L$. In SI Section 2, we explicitly calculate input-output mutual information, $I(R;TF)$, for networks of varying depth at both binding scenarios. We also examined the effect of input distributions on mutual information (see SI Section 2 for details).  As expected, the mutual information is zero when input is completely certain, \emph{viz}. $q=0,1$.  When binding is tight (i.e., $\beta\mu \lesssim -1$), the optimal input distribution $q^\star$ that maximizes the mutual information is $q^\star= 0.5$ -- the input distribution with highest entropy (see SI Section 2 Figure~\ref{fig:SI-MI_lin_max}). Surprisingly, for weak binding we find numerically and analytically that $q^\star \le 0.5$ (see SI Section 2 Figure~\ref{fig:SI-linear_chain_pd_sol}). 

To summarize, we have found that the binding affinity of interactions can be directly related to the information transmission and the signal-to-noise ratio. We find that weak binding affinities give rise to noise stemming from thermal fluctuations and that this noise can always be reduced by increasing binding affinities and making binding more deterministic.

\subsection*{Noise due to non-specific PPIs mediates the ``information-speed'' trade-off}

The previous observations are hard to reconcile with the observation that many PTM-recognition domains such as SH2 and SH3 have only moderately strong  binding affinities \cite{ladbury2000searching,ladbury2011energetics}. For this reason, we investigated tradeoffs that arise from having strong PPIs. One common requirement of eukaryotic signaling pathways is that they should be able to quickly respond to changes in the environmental conditions. This led us to ask how the strength of PPIs affects kinetics. Stronger binding affinities make it harder for proteins to disassociate, suggesting that there maybe a trade-off between reducing noise and responding quickly in the biophysics of PPIs.

\begin{figure*}
	\centering
	\includegraphics[width=\linewidth]{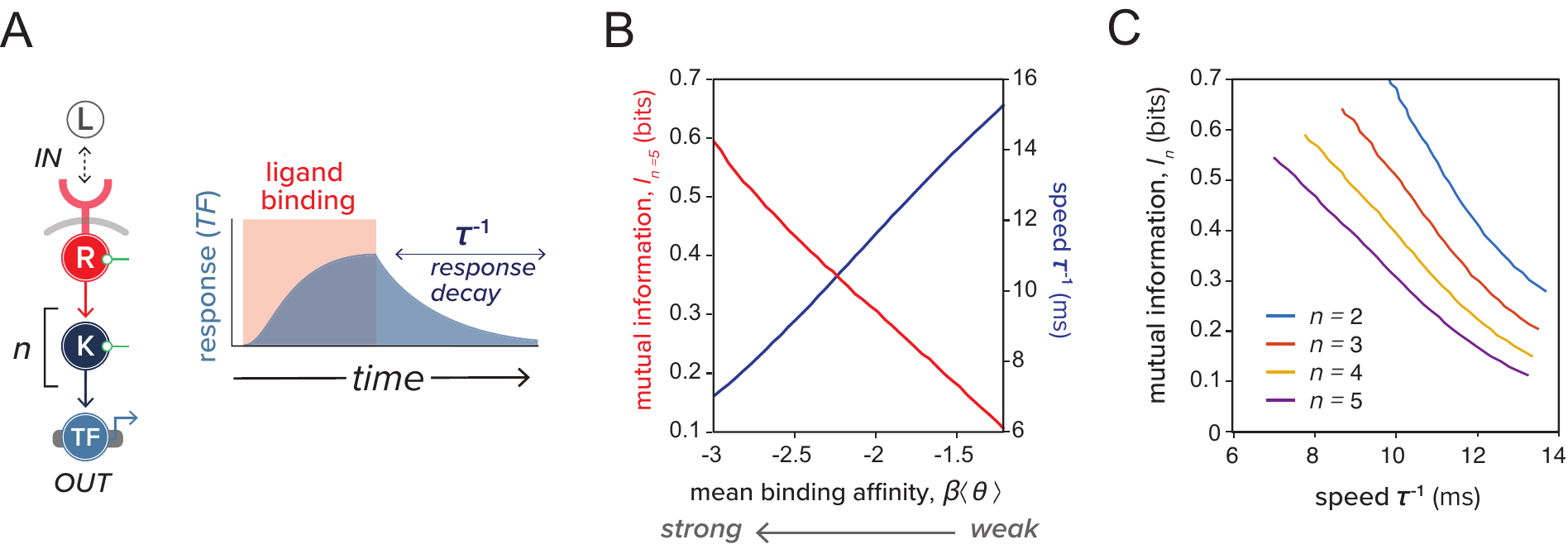}
	\caption{Specificity in PPIs mediates the information-speed tradeoff. ({\it A}) A non-zero constant input (ligand binding of duration indicated in red) is administered to the network shown on the left. This signal turns on the output (TF) to its steady-state before switching off. The speed of response is defined as $\tau^{-1}$, where $\tau$ is the time for output TF to reach a new steady-state after the input is turned off (indicated as response decay in blue).  ({\it B}) Mutual information and response speed as a functions of mean binding affinity $\beta \langle\theta\rangle$ for the network shown in {\it A} with $n=5$. ({\it C}) Mutual information versus response speed as $\beta \langle\theta\rangle$ is varied. Different colors correspond to networks of different depth $n$. }
	\label{fig:tradeoff}
\end{figure*}

To test these ideas, we `translated' our thermodynamic model for the cascade studied in Figure~\ref{fig:tradeoff}A into a kinetic model (see SI Section 3 for details). Note that the thermodynamic model presented in Fig.~\ref{fig:network-model}D can be explicitly derived from the kinetic formulation. Here we invoke this duality to investigate both the signaling dynamics through kinetic formulation as well as the steady-state information capacity by the thermodynamic calculation presented in the previous section. Based on our thermodynamic framework, we first calculated input-output mutual information, \emph{viz}. $I(R; TF)$ in Fig.~\ref{fig:network-model}C, with BAs drawn from distributions with different means $\langle\theta\rangle$. Due to the interplay between the kinetic and thermodynamic picture which we explicitly derived in SI Section 3, we mapped these mean BAs $\langle\theta\rangle$ to their corresponding kinetic rates. The key idea behind this mapping is that the steady state solution of the kinetic model with these rate constants is equivalent to its probabilistic counterpart in the thermodynamic model presented above. For example, the fraction of phosphorylated PK $i$ at steady state is the same as $\P(K_i=1)$ in the thermodynamic model. The BAs in the thermodynamic picture, $\theta_{i,j}$, is related to the Michaelis constant of kinase $j$ phosphorylation reaction by $i$, $K_m$, via $\theta_{i,j} = k_BT \ln (K_m/X_{i}^{SS})$, where $X_i^{SS}$ is the steady state concentration of phosphorylated kinase $i$.

We  performed simulations to measure dynamic response of the signaling circuit to an abrupt perturbation where the input signal was suddenly removed (see Figure~\ref{fig:tradeoff}A). We characterized the response times by measuring the time $\tau$ it took the output to reach a new steady-state. We repeated this procedure for binding affinities drawn from distributions with different means $\langle\theta\rangle$. In Figure~\ref{fig:tradeoff}B, we plot both mutual information and the response speed, defined as the inverse of the response time $\tau^{-1}$, against $\beta\langle\theta\rangle$. This plot shows that response speed and mutual information change in opposite ways as the binding affinity is decreased. Tight-binding (specific PPIs, more negative $\beta \langle\theta\rangle$) allows the network to transmit more information at the expense of a slower dynamical response (see Figure~\ref{fig:tradeoff}C). 

This ``speed-information'' trade-off can be viewed as a biophysical manifestation of the gain-bandwidth tradeoff \cite{detwiler2000engineering}. Intuitively, tighter binding means that the binding off-rate is fairly small compared to the on-rate which is dictated by diffusion. This implies once proteins are bound through specific interactions, the lifetime of the bound complex is long.

\subsection*{Information loss in signaling `can' be mitigated by cross-talks when inputs are correlated}

\begin{figure*}
	\centering
	\includegraphics[width=0.7\linewidth]{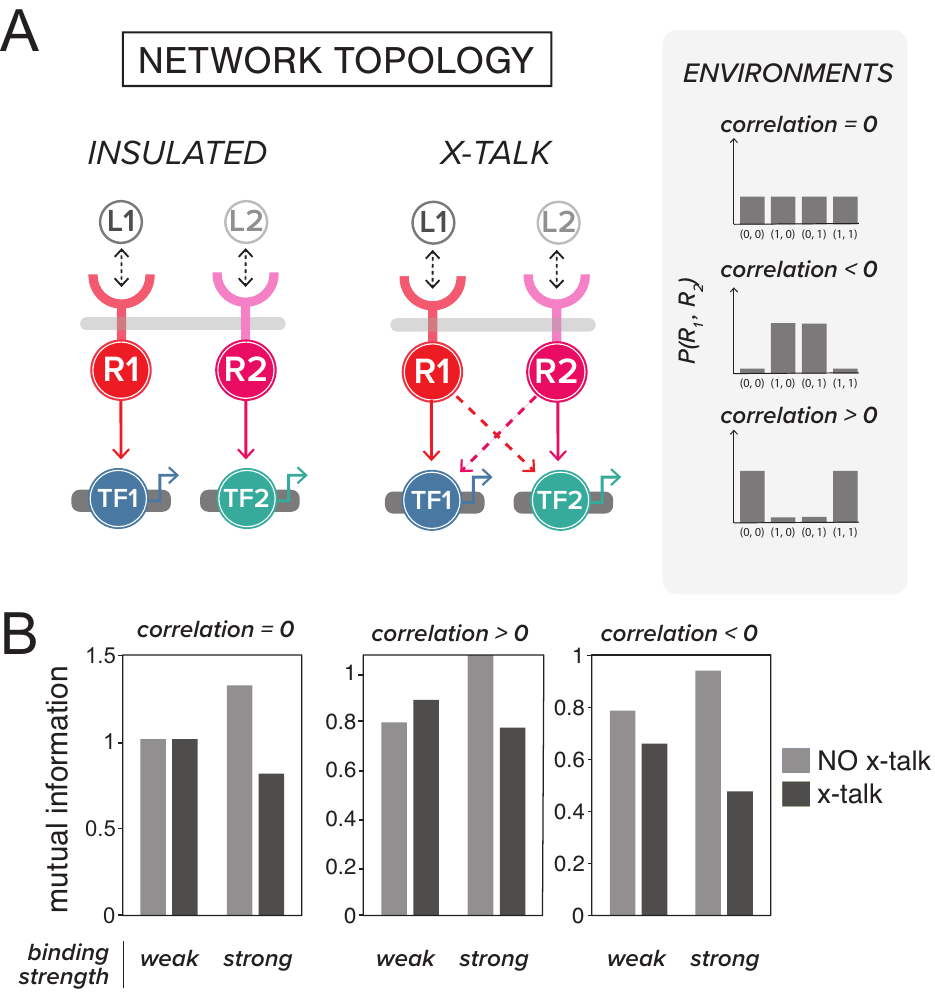}
	\caption{Effect of cross-talks between pathways on information capacity. ({\it A}) Schematic of insulated and cross-talked networks are shown.  Dashed connections represent cross-talks between pathways.  Three environmental conditions that differ in terms of the correlation between the two signals they provide to the network (left) are illustrated: inputs with zero, negative, and positive correlation (see \emph{Materials and Methods} for details). ({\it B}) Mutual information between inputs $(R_1, R_2)$ and outputs (two TFs) is calculated for different input correlations, different mean binding affinities, and  different cross-talk levels. Columns are arranged based on the sign of correlation, bars are grouped according to the strength of binding affinities for protein-protein interactions (weak/tight-binding corresponds to $\beta\langle\theta\rangle = -5/-1$), and colors indicate the presence or absence of cross-talks. For networks with (without) crosstalk, the binding affinity of the cross-talk interactions (i.e. dashed lines in the cross-talked networks shown in {\it A}) is set to $\beta\langle\eta\rangle = -5/0$ (see SI Section 5 for more details.) }
	\label{fig:xtalk}
\end{figure*}

Thus far, we have considered discreet, linearly connected pathways with a single input and output. However, native eukaryotic signaling networks are highly interconnected, with multiple inputs and outputs that cross-talk through PPIs. For this reason, we wanted to better understand how information transduction capacity in multi-input, multi-output (MIMO) networks depended on both the strength of PPIs and the structure of the input signal (i.e. the correlation between inputs). To do so, we studied two parallel pathways, each consisting of an input receptor kinase $R$ and output $TF$ (see Figure~\ref{fig:xtalk}A). In this scheme, cross-talk refers to interactions where proteins in one pathway activate those in the other (i.e., dashed lines in Figure~\ref{fig:xtalk}A). We varied the binding affinity and correlation between two inputs, $R_1$ and $R_2$ -- defined as the connected correlation function (covariance) between the inputs $c\equiv \langle R_1R_2\rangle -\langle R_1\rangle \langle R_2\rangle$ with $\langle\cdot\rangle$ indicating an average over the joint input distribution $\P(R_1, R_2)$ -- and calculated the mutual information, $I(\{R_1, R_2\}; \{TF_1, TF_2\})$ between all the inputs outputs, (see Figure~\ref{fig:xtalk}B for examples).  We found that, regardless of the degree of correlation between inputs, pathway cross-talk is always detrimental to information transmission when noise from non-specific binding is small (i.e., tight-binding).  However, for weak binding and positively correlated inputs, cross-talk can confer a slight benefit, actually increasing information transmission (see Figure~\ref{fig:xtalk}C and SI Section 6 Figure~\ref{fig:SI-2by2stat} for full statistics under the distribution of correlations). This can be rationalized by noting that cross-talk allows cells to reduce noise by "averaging" the two input signals. This averaging is of course only possible if the signals are correlated and contain redundant information.

Our results show that while inter-pathway cross-talk usually degrades information, it may actually provide a benefit when input signals are correlated by reducing noise due to weak PPIs. Simulations on larger pathways confirm this qualitative trend (though it becomes more difficult to define cross-talk for more complex circuits). Finally, we note that our presentation has been limited to the case where cross-talk involves cross-activation between pathways (i.e., pathway 1 activates the pathway 2 intermediate and vice versa). This is reasonable since we have restricted ourselves to considering networks consisting of kinases and some background phosphatase activity. If instead, we had allowed for cross-inhibition between pathways (i.e., pathway 1 inhibits the pathway 2 intermediate and vice versa), information capacity would be slightly increased for negatively correlated signals (results not shown) and diminished for correlated inputs.

\subsection*{Information maximization for complex multi-input, multi-output circuits} 

System-wide studies of phosphorylation-based signaling networks have revealed underlying PPI networks to be highly interconnected \cite{levy2010signaling,breitkreutz2010global}. Here we asked how interconnectivity within signaling network can affect its information capacity. To explore this question, we developed a new algorithm we dub "InfoMax", which identifies the binding affinities and protein concentrations that maximize information transmission for a given network topology. InfoMax, which stands for information maximization, begins with an initial random guess of binding affinities. It then utilizes the thermodynamic framework we developed to calculate the input-output mutual information using these affinities. Optimization is then performed on these affinities to maximize mutual information. Since the explicit functional dependence of mutual information on binding affinity is known (c.f. Eq.\eqref{eq:MIdef} and Figure~\ref{fig:network-model}D), this procedure can be done through a combination of analytic and iterative schemes. To make our approach more generalizable and agnostic to topology, we opted to use simulated annealing to conduct optimization (see Algorithm~\ref{euclid} in \emph{Materials and Methods}). A Python implementation of InfoMax is freely available at the author’s Github repository: \url{https://github.com/chinghao0703/InfomaxDesign}.

\begin{figure*}
	\centering
	\includegraphics[width=0.7\linewidth]{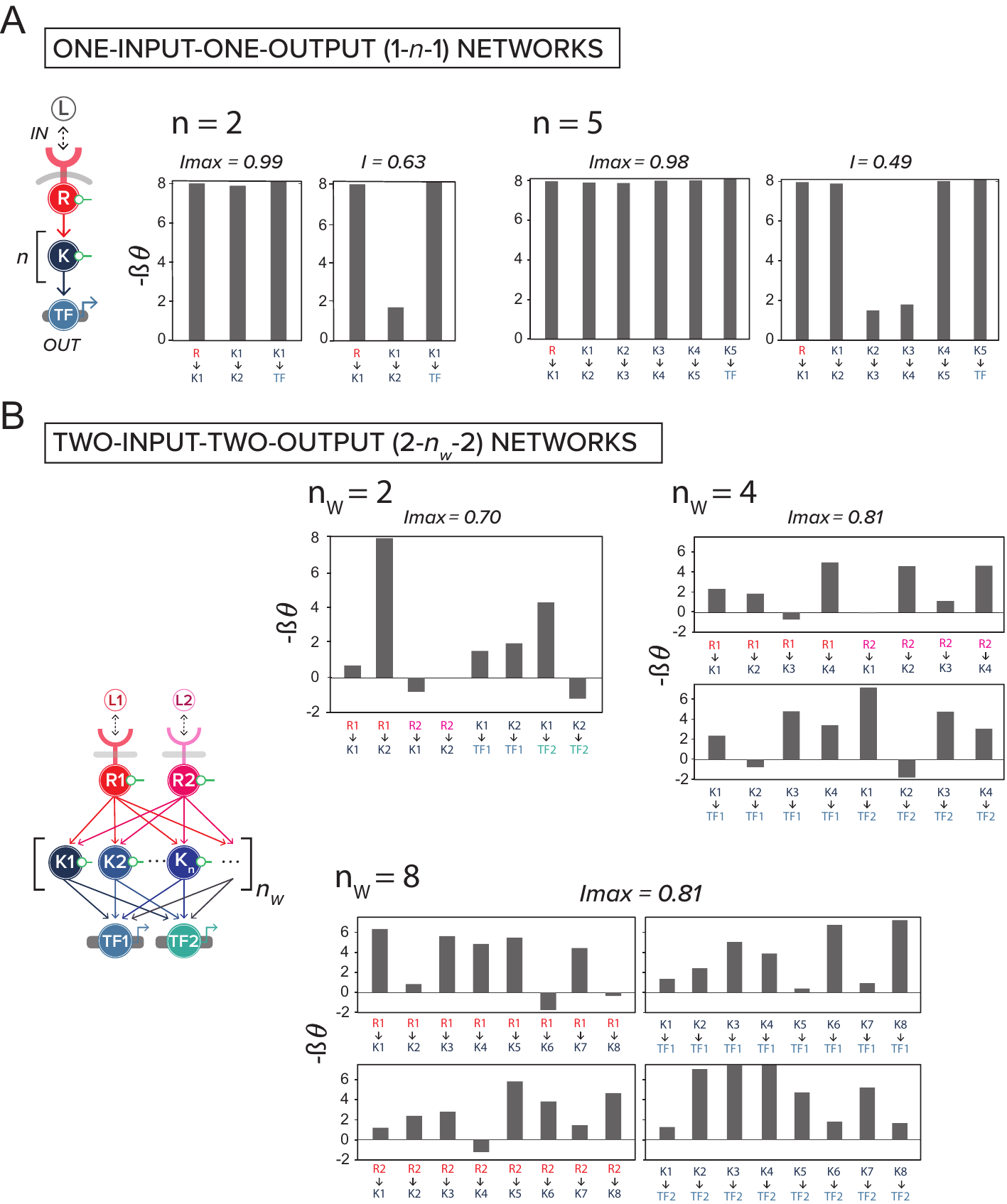}
	\caption{InfoMax design finds the PPIs that maximize information transmission. ({\it A}) InfoMax is applied to networks with one input, $n$ layers of kinase, and one output. For $n=2$ and 5, the bar graph on the left shows the binding affinities that give the maximum mutual information (indicated as $I_{max}$ on top), as opposed to a non-optimal solution with mutual information $I < I_{max}$ shown on the right, all measured in bits. Bars indicate binding affinities between proteins. For example, bar labeled as $R\rightarrow K_1$ is the binding affinity of receptor kinase $R$ to kinase $K_1$. ({\it B}) InfoMax applied to 2-$n_w$-2 networks. This nomenclature refers to all-to-all connected networks with two nodes in the input layer, $n_w$ in the hidden, and two at the output layer. Binding affinities of all networks are optimized to achieve maximum mutual information using simulated annealing (see \emph{Materials and Methods}). Bar charts show the optimized binding affinities with $I_{max}$ indicated on top. In this panel, networks are subject to inputs with zero correlation. In all panels, $-\beta\theta$ is constrained to be within $[-2,8]$.}
	\label{fig:infomax}
\end{figure*}

In order to test the utility of InfoMax, we constructed a library of one-input-one-output networks where we systematically varied  network depth ($n$ in Figure~\ref{fig:infomax}A) and  two-input-two-output networks where we varied the width  ($n_w$ in Figure~\ref{fig:infomax}B). PPI affinities in these networks were optimized using the InfoMax algorithm, allowing us to identify the PPIs configuration with the highest maximum mutual information, subject to the constraints that BAs are bounded within a given range.  In the one-input-one-output networks, we found that increasing network depth always decreased information transmission. This can be understood by noting that additional signaling layers increase non-specific PPI-mediated noise, without ever increasing the strength of the input signal. This observation is a manifestation of the data-processing inequality (DPI), which states that information is never gained by addition more layers when transmitting across noisy channels \cite{cover2012elements,kinney2014equitability}( see Figure~\ref{fig:infomax}A and SI Section 4 Figure~\ref{fig:SI-DPI}). For the optimal solution found, mutual information saturates around 1 bit. As expected, introducing perturbations to this optimal solution by using a sup-optimal binding affinity  at an intermediate layer in the kinase cascade substantially diminish information transmission (see Figure~\ref{fig:infomax}A). 

In contrast, for the two-input-two-output networks we found that increasing the width of the intermediate network can increase information transmission modestly for small widths. As seen in Figure~\ref{fig:infomax}B, these gains quickly saturate after the network reaches the 2-4-2 topology ($n_w=4$). This suggests that modestly widening networks can alleviate bottlenecks in information transmission by reducing noise from weak PPIs. Interestingly, InfoMax also reveals that the optimal PPI-design strategy for more complicated networks can be quite different from the one-input-one-output case where it is always advantageous to have tight-binding between all proteins. In general, we find that the binding affinities that maximize information transmission for multi-input-multi-output networks take on a wide variety of values (see Figure~\ref{fig:infomax}B). 

\section*{Discussions}

The ability of cells to reliably transduce environmental signals is critical for their survival, growth, and proliferation. In this article, we developed a theoretical framework for relating the biophysics of post-translational modifications (PTMs) and protein-protein interactions (PPIs) to information processing in eukaryotic signaling networks. We showed that PPIs with moderate binding affinities necessarily result in thermal noise that limits information transmission within a signaling pathway. While noise can be reduced by increasing binding affinities, this comes with the expense of sluggish dynamic responses, highlighting a fundamental trade-off between information and signaling pathway response dynamics. Although extensive pathway cross-talk is relatively common in signaling networks, we found that it confers little or no advantage to a signaling networks information capacity. 

Our results are consistent with other theoretical works that implicate noise as a major source of information transmission error in signaling \cite{detwiler2000engineering, tkavcik2009optimizing, mehta2009information, walczak2010optimizing, tostevin2009mutual, cheong2011information, brennan2012information}. What is novel about in this work is the ability to directly trace the origin of noise in eukaryotic signaling networks to the strength of PTM-mediated protein-protein interactions. Our results on cross talk also agree with those obtained in \cite{tareen2018modeling} that cross-talks degrades information for channels where input noise can be neglected and inputs are uncorrelated. In addition, our information-theoretic analysis reveals the disadvantages of a deep signaling network, particularly in the face of high non-specific binding (see SI Section 4 Figure~\ref{fig:SI-DPI}). This is consistent with previous work on MAP kinase cascade \cite{detwiler2000engineering}, where the authors argued that maintaining fast response times requires a smaller number of steps with a higher gain per node in order to overcome molecular shot noise. Our simulations also show that information transmission quickly degrades for depths larger than three, which potentially explains the ubiquity of MAP-kinase cascades.

Our work has interesting implications for both natural and synthetic circuits.   A recent study of the  kinase-phosphatase interaction network in budding yeast identified 1844 interactions in budding yeast. Somewhat surprisingly, the binding affinities of many of the identified interactions fell into a narrow affinity window\cite{breitkreutz2010global}. Binding affinity clustering was particularly pronounced for the kinase/phosphatase catalytic domains that mediate phosphorylation-dependent binding\cite{mok2010deciphering}. Our work on the information-speed tradeoff outlined above suggests such an optimized affinity range could be a common feature of networks that need to transmit signals reliably yet quickly in response to noisy environments \cite{ladbury2012noise}.

Another intriguing observation from the yeast kinase-phosphatase interactions is the existence of extensive cross-talk between signaling pathways \cite{breitkreutz2010global} that the authors describe a `collaborative network of interactions' -- a topology that suggests a distributed cellular decision-making strategy \cite{levy2010signaling}. In this article, we show that cross-talk, while unlikely to increase information transmission, is also not particularly detrimental for signaling. Thus, widespread experimental observations of cross-talk in yeast signaling networks likely has an alternative origin. An intriguing hypothesis is that cross-talk arises because of evolutionary selection for signaling robustness\cite{levy2010signaling}. Distributing information processing tasks to many interacting proteins may allow cells to maintaining reliable information transmission even when proteins are deleted or modified.

Our study is directly inspired by synthetic biology, where a long-standing engineering goal is to create cell-based therapies by reprogramming the way in which cells interact with their environment\cite{fischbach2013cell}. Creating synthetic kinase-based signaling circuitry that enables user-customized sense and respond function will necessarily involve information processing considerations, and may favor circuit designs that maximize mutual information between receptor-mediated input and transcriptional output. The potential design space for signaling circuits is vast-unlike genetic circuits, signaling circuits consists of freely diffusible molecular components and thus possess many more tunable parameters that have to be accounted for during design, including circuit topology, intracellular species concentrations, lifetimes, interaction affinities, and intrinsic catalytic rates \cite{bashor2018understanding}. Conclusions from our work suggest some general rules that could be used to constrain the search for productive circuit configurations. For example, focusing on engineering high interaction specificity for parts that mediate PTM-mediated PPIs could potentially mitigate noise, while using Infomax could be used to maximize the information capacity for a given circuit architecture.

\section*{Materials and Methods}
\subsection*{Optimizing mutual information with simulated annealing} Let $G$ be the network given and $\{\theta^{(0)}\}$ be the initial BAs (parameters) which we sample uniformly from $[-8, 2]$. For each time step $t$ $(t=0,\cdots, T=10^6)$, we either add to each elements of $\{\theta^{(t)}\}$ a fixed finite amount $\pm\delta\theta=\pm 10^{-4}$ or leave it un-perturbed, completely at random. The perturbed parameter $\{\theta^{(t)^\star}\}$is accepted with probability $p= \min \{1, \exp\{\alpha [I(\{\theta^{(t)^\star}\})-I(\{\theta^{(t)}\})]\}\}$, where $\alpha = \log t$. If accepted, set $\{\theta^{(t+1)}\}\leftarrow \{\theta^{(t)^\star}\}$; otherwise, set $\{\theta^{(t+1)}\}\leftarrow \{\theta^{(t)}\}$. This procedure continues until any element in $\{\theta^{(t)}\}$ falls beyond $[-8, 2]$ or $t=T$, whichever happens earlier. Python code for such implementation is available at the author's Github repository: \url{https://github.com/chinghao0703/InfomaxDesign}. In Figure~\ref{fig:infomax}, we perform 100 simulated annealing routines with un-correlated inputs and report the realization that gives maximum mutual information denoted as $I_\text{max}$. The pseudo-code of the InfoMax procedure is given in Algorithm~\ref{euclid}.

\begin{algorithm}
	\caption{InfoMax: maximizing mutual information with respect to BAs}\label{euclid}
	\begin{algorithmic}[1]\label{algInfoMax}
		\Require Binding affinity (BA) range  $S\leftarrow [-8.0, 2.0]$, perturbation $\delta\theta \leftarrow 10^{-4}$, termination threshold  $T \leftarrow 10^6$, and function $I(\cdot)$ that computes the input-output mutual information of a given signaling network $G$ with BA $\{\theta\}$. 
		\Procedure{InfoMax}{$G$}
		\State $t\gets 0$
		\State Draw BA uniformly from $S$: $\{\thet\}\gets U(S)$ 
		\While {$t\le T$ \texttt{ AND } $\{\thet\}\in S$}
		\State{Draw $a$ uniformly from $\{\-1, 0, -1\}$ and update BA $\{\thets\}\leftarrow \{\thet + a\cdot\delta\theta$\} }
		\State{Compute $p\leftarrow \min \{1, \exp\{\alpha [I(\{\theta^{(t)^\star}\})-I(\{\theta^{(t)}\})]\}\}$, where $\alpha = \log t$}
		\State{Draw $b$ uniformly from $[0,1]$}
		
		\If {$b \le p$} 
		\State Accept $\{\thetp\} \leftarrow \{\thets\}$ 
		\Else 
		\State $\{\thetp\} \leftarrow \{\thet\}$ 
		\EndIf
		\State $t+1 \gets t$
		\EndWhile

		\Return $\{\thet\}$
		\EndProcedure
	\end{algorithmic}
\end{algorithm}


\subsection*{Calculating correlation between inputs}
Here we consider two-input-two-output all-to-all connected network. As before, let $x_1, x_2$ be the inputs while $x_3, x_4$ be the outputs. Let's also define the connected correlation functions for the input as:
\be
c=\E [x_1x_2] - \E[x_1]\E[x_2],
\ee 
where $\E[\cdots] $ is taken with respect to the joint distribution $\P(x_1,x_2)$ given by
\begin{center}
	\begin{tabular}{|c| c|} 
		\hline
		$(x_1,x_2)$ & $\P(x_1,x_2)$ \\ [0.5ex] 
		\hline\hline
		$(0,0)$ & $q_1$ \\
		\hline
		$(1,0)$ & $q_2$ \\
		\hline
		$(0,1)$ & $q_3$ \\
		\hline $(1,1)$ & $1-q_1-q_2-q_3$ \ \\ [1ex] 
		\hline
	\end{tabular}
\end{center}
From this, the connected correlation function reads
\bea\label{eq:corr}
c&=&q_1(1-q_1-q_2-q_3) -q_2q_3\nn\\
&=& \P(0,0)\P(1,1)- \P(1,0)\P(0,1)
\eea

\section*{Acknowledgments}

We thank Amir Bitran and Henry Mattingly for helpful discussions.  This work was also supported by NIH NIGMS grant 1R35GM119461, and by Simons Investigator in the Mathematical Modeling of Living Systems (MMLS) awards to PM. Part of the computations were
carried out on the Boston University Shared Computing Cluster (SCC).

\bibliography{infomax.bib}

\begin{thebibliography}{48}%
\makeatletter
\providecommand \@ifxundefined [1]{%
 \@ifx{#1\undefined}
}%
\providecommand \@ifnum [1]{%
 \ifnum #1\expandafter \@firstoftwo
 \else \expandafter \@secondoftwo
 \fi
}%
\providecommand \@ifx [1]{%
 \ifx #1\expandafter \@firstoftwo
 \else \expandafter \@secondoftwo
 \fi
}%
\providecommand \natexlab [1]{#1}%
\providecommand \enquote  [1]{``#1''}%
\providecommand \bibnamefont  [1]{#1}%
\providecommand \bibfnamefont [1]{#1}%
\providecommand \citenamefont [1]{#1}%
\providecommand \href@noop [0]{\@secondoftwo}%
\providecommand \href [0]{\begingroup \@sanitize@url \@href}%
\providecommand \@href[1]{\@@startlink{#1}\@@href}%
\providecommand \@@href[1]{\endgroup#1\@@endlink}%
\providecommand \@sanitize@url [0]{\catcode `\\12\catcode `\$12\catcode
  `\&12\catcode `\#12\catcode `\^12\catcode `\_12\catcode `\%12\relax}%
\providecommand \@@startlink[1]{}%
\providecommand \@@endlink[0]{}%
\providecommand \url  [0]{\begingroup\@sanitize@url \@url }%
\providecommand \@url [1]{\endgroup\@href {#1}{\urlprefix }}%
\providecommand \urlprefix  [0]{URL }%
\providecommand \Eprint [0]{\href }%
\providecommand \doibase [0]{http://dx.doi.org/}%
\providecommand \selectlanguage [0]{\@gobble}%
\providecommand \bibinfo  [0]{\@secondoftwo}%
\providecommand \bibfield  [0]{\@secondoftwo}%
\providecommand \translation [1]{[#1]}%
\providecommand \BibitemOpen [0]{}%
\providecommand \bibitemStop [0]{}%
\providecommand \bibitemNoStop [0]{.\EOS\space}%
\providecommand \EOS [0]{\spacefactor3000\relax}%
\providecommand \BibitemShut  [1]{\csname bibitem#1\endcsname}%
\let\auto@bib@innerbib\@empty
\bibitem [{\citenamefont {Barabasi}\ and\ \citenamefont
  {Oltvai}(2004)}]{barabasi2004network}%
  \BibitemOpen
  \bibfield  {author} {\bibinfo {author} {\bibfnamefont {A.-L.}\ \bibnamefont
  {Barabasi}}\ and\ \bibinfo {author} {\bibfnamefont {Z.~N.}\ \bibnamefont
  {Oltvai}},\ }\href@noop {} {\bibfield  {journal} {\bibinfo  {journal} {Nature
  reviews genetics}\ }\textbf {\bibinfo {volume} {5}},\ \bibinfo {pages} {101}
  (\bibinfo {year} {2004})}\BibitemShut {NoStop}%
\bibitem [{\citenamefont {Blais}\ and\ \citenamefont
  {Dynlacht}(2005)}]{blais2005constructing}%
  \BibitemOpen
  \bibfield  {author} {\bibinfo {author} {\bibfnamefont {A.}~\bibnamefont
  {Blais}}\ and\ \bibinfo {author} {\bibfnamefont {B.~D.}\ \bibnamefont
  {Dynlacht}},\ }\href@noop {} {\bibfield  {journal} {\bibinfo  {journal}
  {Genes \& development}\ }\textbf {\bibinfo {volume} {19}},\ \bibinfo {pages}
  {1499} (\bibinfo {year} {2005})}\BibitemShut {NoStop}%
\bibitem [{\citenamefont {MacArthur}\ \emph {et~al.}(2009)\citenamefont
  {MacArthur}, \citenamefont {Ma'ayan},\ and\ \citenamefont
  {Lemischka}}]{macarthur2009systems}%
  \BibitemOpen
  \bibfield  {author} {\bibinfo {author} {\bibfnamefont {B.~D.}\ \bibnamefont
  {MacArthur}}, \bibinfo {author} {\bibfnamefont {A.}~\bibnamefont {Ma'ayan}},
  \ and\ \bibinfo {author} {\bibfnamefont {I.~R.}\ \bibnamefont {Lemischka}},\
  }\href@noop {} {\bibfield  {journal} {\bibinfo  {journal} {Nature Reviews
  Molecular Cell Biology}\ }\textbf {\bibinfo {volume} {10}},\ \bibinfo {pages}
  {672} (\bibinfo {year} {2009})}\BibitemShut {NoStop}%
\bibitem [{\citenamefont {Martello}\ and\ \citenamefont
  {Smith}(2014)}]{martello2014nature}%
  \BibitemOpen
  \bibfield  {author} {\bibinfo {author} {\bibfnamefont {G.}~\bibnamefont
  {Martello}}\ and\ \bibinfo {author} {\bibfnamefont {A.}~\bibnamefont
  {Smith}},\ }\href@noop {} {\bibfield  {journal} {\bibinfo  {journal} {Annual
  review of cell and developmental biology}\ }\textbf {\bibinfo {volume}
  {30}},\ \bibinfo {pages} {647} (\bibinfo {year} {2014})}\BibitemShut
  {NoStop}%
\bibitem [{\citenamefont {Seet}\ \emph {et~al.}(2006)\citenamefont {Seet},
  \citenamefont {Dikic}, \citenamefont {Zhou},\ and\ \citenamefont
  {Pawson}}]{seet2006reading}%
  \BibitemOpen
  \bibfield  {author} {\bibinfo {author} {\bibfnamefont {B.~T.}\ \bibnamefont
  {Seet}}, \bibinfo {author} {\bibfnamefont {I.}~\bibnamefont {Dikic}},
  \bibinfo {author} {\bibfnamefont {M.-M.}\ \bibnamefont {Zhou}}, \ and\
  \bibinfo {author} {\bibfnamefont {T.}~\bibnamefont {Pawson}},\ }\href@noop {}
  {\bibfield  {journal} {\bibinfo  {journal} {Nature reviews Molecular cell
  biology}\ }\textbf {\bibinfo {volume} {7}},\ \bibinfo {pages} {473} (\bibinfo
  {year} {2006})}\BibitemShut {NoStop}%
\bibitem [{\citenamefont {Scott}\ and\ \citenamefont
  {Pawson}(2009)}]{scott2009cell}%
  \BibitemOpen
  \bibfield  {author} {\bibinfo {author} {\bibfnamefont {J.~D.}\ \bibnamefont
  {Scott}}\ and\ \bibinfo {author} {\bibfnamefont {T.}~\bibnamefont {Pawson}},\
  }\href@noop {} {\bibfield  {journal} {\bibinfo  {journal} {Science}\ }\textbf
  {\bibinfo {volume} {326}},\ \bibinfo {pages} {1220} (\bibinfo {year}
  {2009})}\BibitemShut {NoStop}%
\bibitem [{\citenamefont {Lim}\ \emph {et~al.}(2014)\citenamefont {Lim},
  \citenamefont {Mayer},\ and\ \citenamefont {Pawson}}]{lim2014cell}%
  \BibitemOpen
  \bibfield  {author} {\bibinfo {author} {\bibfnamefont {W.}~\bibnamefont
  {Lim}}, \bibinfo {author} {\bibfnamefont {B.}~\bibnamefont {Mayer}}, \ and\
  \bibinfo {author} {\bibfnamefont {T.}~\bibnamefont {Pawson}},\ }\href@noop {}
  {\emph {\bibinfo {title} {Cell signaling: principles and mechanisms}}}\
  (\bibinfo  {publisher} {Taylor \& Francis},\ \bibinfo {year}
  {2014})\BibitemShut {NoStop}%
\bibitem [{\citenamefont {Angers}\ and\ \citenamefont
  {Moon}(2009)}]{angers2009proximal}%
  \BibitemOpen
  \bibfield  {author} {\bibinfo {author} {\bibfnamefont {S.}~\bibnamefont
  {Angers}}\ and\ \bibinfo {author} {\bibfnamefont {R.~T.}\ \bibnamefont
  {Moon}},\ }\href@noop {} {\bibfield  {journal} {\bibinfo  {journal} {Nature
  reviews Molecular cell biology}\ }\textbf {\bibinfo {volume} {10}},\ \bibinfo
  {pages} {468} (\bibinfo {year} {2009})}\BibitemShut {NoStop}%
\bibitem [{\citenamefont {MacDonald}\ \emph {et~al.}(2009)\citenamefont
  {MacDonald}, \citenamefont {Tamai},\ and\ \citenamefont
  {He}}]{macdonald2009wnt}%
  \BibitemOpen
  \bibfield  {author} {\bibinfo {author} {\bibfnamefont {B.~T.}\ \bibnamefont
  {MacDonald}}, \bibinfo {author} {\bibfnamefont {K.}~\bibnamefont {Tamai}}, \
  and\ \bibinfo {author} {\bibfnamefont {X.}~\bibnamefont {He}},\ }\href@noop
  {} {\bibfield  {journal} {\bibinfo  {journal} {Developmental cell}\ }\textbf
  {\bibinfo {volume} {17}},\ \bibinfo {pages} {9} (\bibinfo {year}
  {2009})}\BibitemShut {NoStop}%
\bibitem [{\citenamefont {Massagu{\'e}}(2012)}]{massague2012tgfbeta}%
  \BibitemOpen
  \bibfield  {author} {\bibinfo {author} {\bibfnamefont {J.}~\bibnamefont
  {Massagu{\'e}}},\ }\href@noop {} {\bibfield  {journal} {\bibinfo  {journal}
  {Nature reviews Molecular cell biology}\ }\textbf {\bibinfo {volume} {13}},\
  \bibinfo {pages} {616} (\bibinfo {year} {2012})}\BibitemShut {NoStop}%
\bibitem [{\citenamefont {Anastas}\ and\ \citenamefont
  {Moon}(2013)}]{anastas2013wnt}%
  \BibitemOpen
  \bibfield  {author} {\bibinfo {author} {\bibfnamefont {J.~N.}\ \bibnamefont
  {Anastas}}\ and\ \bibinfo {author} {\bibfnamefont {R.~T.}\ \bibnamefont
  {Moon}},\ }\href@noop {} {\bibfield  {journal} {\bibinfo  {journal} {Nature
  Reviews Cancer}\ }\textbf {\bibinfo {volume} {13}},\ \bibinfo {pages} {11}
  (\bibinfo {year} {2013})}\BibitemShut {NoStop}%
\bibitem [{\citenamefont {Moustakas}\ and\ \citenamefont
  {Heldin}(2014)}]{moustakas2014tgfbeta}%
  \BibitemOpen
  \bibfield  {author} {\bibinfo {author} {\bibfnamefont {A.}~\bibnamefont
  {Moustakas}}\ and\ \bibinfo {author} {\bibfnamefont {P.}~\bibnamefont
  {Heldin}},\ }\href@noop {} {\bibfield  {journal} {\bibinfo  {journal}
  {Biochimica et Biophysica Acta (BBA)-General Subjects}\ }\textbf {\bibinfo
  {volume} {1840}},\ \bibinfo {pages} {2621} (\bibinfo {year}
  {2014})}\BibitemShut {NoStop}%
\bibitem [{\citenamefont {Deribe}\ \emph {et~al.}(2010)\citenamefont {Deribe},
  \citenamefont {Pawson},\ and\ \citenamefont {Dikic}}]{deribe2010post}%
  \BibitemOpen
  \bibfield  {author} {\bibinfo {author} {\bibfnamefont {Y.~L.}\ \bibnamefont
  {Deribe}}, \bibinfo {author} {\bibfnamefont {T.}~\bibnamefont {Pawson}}, \
  and\ \bibinfo {author} {\bibfnamefont {I.}~\bibnamefont {Dikic}},\
  }\href@noop {} {\bibfield  {journal} {\bibinfo  {journal} {Nature structural
  \& molecular biology}\ }\textbf {\bibinfo {volume} {17}},\ \bibinfo {pages}
  {666} (\bibinfo {year} {2010})}\BibitemShut {NoStop}%
\bibitem [{\citenamefont {Ackers}\ \emph {et~al.}(1982)\citenamefont {Ackers},
  \citenamefont {Johnson},\ and\ \citenamefont
  {Shea}}]{ackers1982quantitative}%
  \BibitemOpen
  \bibfield  {author} {\bibinfo {author} {\bibfnamefont {G.~K.}\ \bibnamefont
  {Ackers}}, \bibinfo {author} {\bibfnamefont {A.~D.}\ \bibnamefont {Johnson}},
  \ and\ \bibinfo {author} {\bibfnamefont {M.~A.}\ \bibnamefont {Shea}},\
  }\href@noop {} {\bibfield  {journal} {\bibinfo  {journal} {Proceedings of the
  National Academy of Sciences}\ }\textbf {\bibinfo {volume} {79}},\ \bibinfo
  {pages} {1129} (\bibinfo {year} {1982})}\BibitemShut {NoStop}%
\bibitem [{\citenamefont {Hill}(2013)}]{hill2013cooperativity}%
  \BibitemOpen
  \bibfield  {author} {\bibinfo {author} {\bibfnamefont {T.~L.}\ \bibnamefont
  {Hill}},\ }\href@noop {} {\emph {\bibinfo {title} {Cooperativity theory in
  biochemistry: steady-state and equilibrium systems}}}\ (\bibinfo  {publisher}
  {Springer Science \& Business Media},\ \bibinfo {year} {2013})\BibitemShut
  {NoStop}%
\bibitem [{\citenamefont {Weinert}\ \emph {et~al.}(2014)\citenamefont
  {Weinert}, \citenamefont {Brewster}, \citenamefont {Rydenfelt}, \citenamefont
  {Phillips},\ and\ \citenamefont {Kegel}}]{weinert2014scaling}%
  \BibitemOpen
  \bibfield  {author} {\bibinfo {author} {\bibfnamefont {F.~M.}\ \bibnamefont
  {Weinert}}, \bibinfo {author} {\bibfnamefont {R.~C.}\ \bibnamefont
  {Brewster}}, \bibinfo {author} {\bibfnamefont {M.}~\bibnamefont {Rydenfelt}},
  \bibinfo {author} {\bibfnamefont {R.}~\bibnamefont {Phillips}}, \ and\
  \bibinfo {author} {\bibfnamefont {W.~K.}\ \bibnamefont {Kegel}},\ }\href@noop
  {} {\bibfield  {journal} {\bibinfo  {journal} {Physical review letters}\
  }\textbf {\bibinfo {volume} {113}},\ \bibinfo {pages} {258101} (\bibinfo
  {year} {2014})}\BibitemShut {NoStop}%
\bibitem [{\citenamefont {Bintu}\ \emph {et~al.}(2005)\citenamefont {Bintu},
  \citenamefont {Buchler}, \citenamefont {Garcia}, \citenamefont {Gerland},
  \citenamefont {Hwa}, \citenamefont {Kondev},\ and\ \citenamefont
  {Phillips}}]{bintu2005transcriptional}%
  \BibitemOpen
  \bibfield  {author} {\bibinfo {author} {\bibfnamefont {L.}~\bibnamefont
  {Bintu}}, \bibinfo {author} {\bibfnamefont {N.~E.}\ \bibnamefont {Buchler}},
  \bibinfo {author} {\bibfnamefont {H.~G.}\ \bibnamefont {Garcia}}, \bibinfo
  {author} {\bibfnamefont {U.}~\bibnamefont {Gerland}}, \bibinfo {author}
  {\bibfnamefont {T.}~\bibnamefont {Hwa}}, \bibinfo {author} {\bibfnamefont
  {J.}~\bibnamefont {Kondev}}, \ and\ \bibinfo {author} {\bibfnamefont
  {R.}~\bibnamefont {Phillips}},\ }\href@noop {} {\bibfield  {journal}
  {\bibinfo  {journal} {Current opinion in genetics \& development}\ }\textbf
  {\bibinfo {volume} {15}},\ \bibinfo {pages} {116} (\bibinfo {year}
  {2005})}\BibitemShut {NoStop}%
\bibitem [{\citenamefont {Kinney}\ \emph {et~al.}(2010)\citenamefont {Kinney},
  \citenamefont {Murugan}, \citenamefont {Callan},\ and\ \citenamefont
  {Cox}}]{kinney2010using}%
  \BibitemOpen
  \bibfield  {author} {\bibinfo {author} {\bibfnamefont {J.~B.}\ \bibnamefont
  {Kinney}}, \bibinfo {author} {\bibfnamefont {A.}~\bibnamefont {Murugan}},
  \bibinfo {author} {\bibfnamefont {C.~G.}\ \bibnamefont {Callan}}, \ and\
  \bibinfo {author} {\bibfnamefont {E.~C.}\ \bibnamefont {Cox}},\ }\href@noop
  {} {\bibfield  {journal} {\bibinfo  {journal} {Proceedings of the National
  Academy of Sciences}\ } (\bibinfo {year} {2010})}\BibitemShut {NoStop}%
\bibitem [{\citenamefont {Garcia}\ \emph {et~al.}(2010)\citenamefont {Garcia},
  \citenamefont {Sanchez}, \citenamefont {Kuhlman}, \citenamefont {Kondev},\
  and\ \citenamefont {Phillips}}]{garcia2010transcription}%
  \BibitemOpen
  \bibfield  {author} {\bibinfo {author} {\bibfnamefont {H.~G.}\ \bibnamefont
  {Garcia}}, \bibinfo {author} {\bibfnamefont {A.}~\bibnamefont {Sanchez}},
  \bibinfo {author} {\bibfnamefont {T.}~\bibnamefont {Kuhlman}}, \bibinfo
  {author} {\bibfnamefont {J.}~\bibnamefont {Kondev}}, \ and\ \bibinfo {author}
  {\bibfnamefont {R.}~\bibnamefont {Phillips}},\ }\href@noop {} {\bibfield
  {journal} {\bibinfo  {journal} {Trends in cell biology}\ }\textbf {\bibinfo
  {volume} {20}},\ \bibinfo {pages} {723} (\bibinfo {year} {2010})}\BibitemShut
  {NoStop}%
\bibitem [{\citenamefont {Shannon}(2001)}]{shannon2001mathematical}%
  \BibitemOpen
  \bibfield  {author} {\bibinfo {author} {\bibfnamefont {C.~E.}\ \bibnamefont
  {Shannon}},\ }\href@noop {} {\bibfield  {journal} {\bibinfo  {journal} {ACM
  SIGMOBILE mobile computing and communications review}\ }\textbf {\bibinfo
  {volume} {5}},\ \bibinfo {pages} {3} (\bibinfo {year} {2001})}\BibitemShut
  {NoStop}%
\bibitem [{\citenamefont {Cover}\ and\ \citenamefont
  {Thomas}(2012)}]{cover2012elements}%
  \BibitemOpen
  \bibfield  {author} {\bibinfo {author} {\bibfnamefont {T.~M.}\ \bibnamefont
  {Cover}}\ and\ \bibinfo {author} {\bibfnamefont {J.~A.}\ \bibnamefont
  {Thomas}},\ }\href@noop {} {\emph {\bibinfo {title} {Elements of information
  theory}}}\ (\bibinfo  {publisher} {John Wiley \& Sons},\ \bibinfo {year}
  {2012})\BibitemShut {NoStop}%
\bibitem [{\citenamefont {Johnson}(1970)}]{johnson1970information}%
  \BibitemOpen
  \bibfield  {author} {\bibinfo {author} {\bibfnamefont {H.~A.}\ \bibnamefont
  {Johnson}},\ }\href@noop {} {\bibfield  {journal} {\bibinfo  {journal}
  {Science}\ }\textbf {\bibinfo {volume} {168}},\ \bibinfo {pages} {1545}
  (\bibinfo {year} {1970})}\BibitemShut {NoStop}%
\bibitem [{\citenamefont {Tka{\v{c}}ik}\ \emph {et~al.}(2009)\citenamefont
  {Tka{\v{c}}ik}, \citenamefont {Walczak},\ and\ \citenamefont
  {Bialek}}]{tkavcik2009optimizing}%
  \BibitemOpen
  \bibfield  {author} {\bibinfo {author} {\bibfnamefont {G.}~\bibnamefont
  {Tka{\v{c}}ik}}, \bibinfo {author} {\bibfnamefont {A.~M.}\ \bibnamefont
  {Walczak}}, \ and\ \bibinfo {author} {\bibfnamefont {W.}~\bibnamefont
  {Bialek}},\ }\href@noop {} {\bibfield  {journal} {\bibinfo  {journal}
  {Physical Review E}\ }\textbf {\bibinfo {volume} {80}},\ \bibinfo {pages}
  {031920} (\bibinfo {year} {2009})}\BibitemShut {NoStop}%
\bibitem [{\citenamefont {Walczak}\ \emph {et~al.}(2010)\citenamefont
  {Walczak}, \citenamefont {Tka{\v{c}}ik},\ and\ \citenamefont
  {Bialek}}]{walczak2010optimizing}%
  \BibitemOpen
  \bibfield  {author} {\bibinfo {author} {\bibfnamefont {A.~M.}\ \bibnamefont
  {Walczak}}, \bibinfo {author} {\bibfnamefont {G.}~\bibnamefont
  {Tka{\v{c}}ik}}, \ and\ \bibinfo {author} {\bibfnamefont {W.}~\bibnamefont
  {Bialek}},\ }\href@noop {} {\bibfield  {journal} {\bibinfo  {journal}
  {Physical Review E}\ }\textbf {\bibinfo {volume} {81}},\ \bibinfo {pages}
  {041905} (\bibinfo {year} {2010})}\BibitemShut {NoStop}%
\bibitem [{\citenamefont {Tka{\v{c}}ik}\ and\ \citenamefont
  {Walczak}(2011)}]{tkavcik2011information}%
  \BibitemOpen
  \bibfield  {author} {\bibinfo {author} {\bibfnamefont {G.}~\bibnamefont
  {Tka{\v{c}}ik}}\ and\ \bibinfo {author} {\bibfnamefont {A.~M.}\ \bibnamefont
  {Walczak}},\ }\href@noop {} {\bibfield  {journal} {\bibinfo  {journal}
  {Journal of Physics: Condensed Matter}\ }\textbf {\bibinfo {volume} {23}},\
  \bibinfo {pages} {153102} (\bibinfo {year} {2011})}\BibitemShut {NoStop}%
\bibitem [{\citenamefont {Granados}\ \emph {et~al.}(2018)\citenamefont
  {Granados}, \citenamefont {Pietsch}, \citenamefont {Cepeda-Humerez},
  \citenamefont {Farquhar}, \citenamefont {Tka{\v{c}}ik},\ and\ \citenamefont
  {Swain}}]{granados2018distributed}%
  \BibitemOpen
  \bibfield  {author} {\bibinfo {author} {\bibfnamefont {A.~A.}\ \bibnamefont
  {Granados}}, \bibinfo {author} {\bibfnamefont {J.~M.}\ \bibnamefont
  {Pietsch}}, \bibinfo {author} {\bibfnamefont {S.~A.}\ \bibnamefont
  {Cepeda-Humerez}}, \bibinfo {author} {\bibfnamefont {I.~L.}\ \bibnamefont
  {Farquhar}}, \bibinfo {author} {\bibfnamefont {G.}~\bibnamefont
  {Tka{\v{c}}ik}}, \ and\ \bibinfo {author} {\bibfnamefont {P.~S.}\
  \bibnamefont {Swain}},\ }\href@noop {} {\bibfield  {journal} {\bibinfo
  {journal} {Proceedings of the National Academy of Sciences}\ }\textbf
  {\bibinfo {volume} {115}},\ \bibinfo {pages} {6088} (\bibinfo {year}
  {2018})}\BibitemShut {NoStop}%
\bibitem [{\citenamefont {Detwiler}\ \emph {et~al.}(2000)\citenamefont
  {Detwiler}, \citenamefont {Ramanathan}, \citenamefont {Sengupta},\ and\
  \citenamefont {Shraiman}}]{detwiler2000engineering}%
  \BibitemOpen
  \bibfield  {author} {\bibinfo {author} {\bibfnamefont {P.~B.}\ \bibnamefont
  {Detwiler}}, \bibinfo {author} {\bibfnamefont {S.}~\bibnamefont
  {Ramanathan}}, \bibinfo {author} {\bibfnamefont {A.}~\bibnamefont
  {Sengupta}}, \ and\ \bibinfo {author} {\bibfnamefont {B.~I.}\ \bibnamefont
  {Shraiman}},\ }\href@noop {} {\bibfield  {journal} {\bibinfo  {journal}
  {Biophysical Journal}\ }\textbf {\bibinfo {volume} {79}},\ \bibinfo {pages}
  {2801} (\bibinfo {year} {2000})}\BibitemShut {NoStop}%
\bibitem [{\citenamefont {Mehta}\ \emph {et~al.}(2009)\citenamefont {Mehta},
  \citenamefont {Goyal}, \citenamefont {Long}, \citenamefont {Bassler},\ and\
  \citenamefont {Wingreen}}]{mehta2009information}%
  \BibitemOpen
  \bibfield  {author} {\bibinfo {author} {\bibfnamefont {P.}~\bibnamefont
  {Mehta}}, \bibinfo {author} {\bibfnamefont {S.}~\bibnamefont {Goyal}},
  \bibinfo {author} {\bibfnamefont {T.}~\bibnamefont {Long}}, \bibinfo {author}
  {\bibfnamefont {B.~L.}\ \bibnamefont {Bassler}}, \ and\ \bibinfo {author}
  {\bibfnamefont {N.~S.}\ \bibnamefont {Wingreen}},\ }\href@noop {} {\bibfield
  {journal} {\bibinfo  {journal} {Molecular systems biology}\ }\textbf
  {\bibinfo {volume} {5}},\ \bibinfo {pages} {325} (\bibinfo {year}
  {2009})}\BibitemShut {NoStop}%
\bibitem [{\citenamefont {Tostevin}\ and\ \citenamefont
  {Ten~Wolde}(2009)}]{tostevin2009mutual}%
  \BibitemOpen
  \bibfield  {author} {\bibinfo {author} {\bibfnamefont {F.}~\bibnamefont
  {Tostevin}}\ and\ \bibinfo {author} {\bibfnamefont {P.~R.}\ \bibnamefont
  {Ten~Wolde}},\ }\href@noop {} {\bibfield  {journal} {\bibinfo  {journal}
  {Physical review letters}\ }\textbf {\bibinfo {volume} {102}},\ \bibinfo
  {pages} {218101} (\bibinfo {year} {2009})}\BibitemShut {NoStop}%
\bibitem [{\citenamefont {Cheong}\ \emph {et~al.}(2011)\citenamefont {Cheong},
  \citenamefont {Rhee}, \citenamefont {Wang}, \citenamefont {Nemenman},\ and\
  \citenamefont {Levchenko}}]{cheong2011information}%
  \BibitemOpen
  \bibfield  {author} {\bibinfo {author} {\bibfnamefont {R.}~\bibnamefont
  {Cheong}}, \bibinfo {author} {\bibfnamefont {A.}~\bibnamefont {Rhee}},
  \bibinfo {author} {\bibfnamefont {C.~J.}\ \bibnamefont {Wang}}, \bibinfo
  {author} {\bibfnamefont {I.}~\bibnamefont {Nemenman}}, \ and\ \bibinfo
  {author} {\bibfnamefont {A.}~\bibnamefont {Levchenko}},\ }\href@noop {}
  {\bibfield  {journal} {\bibinfo  {journal} {science}\ ,\ \bibinfo {pages}
  {1204553}} (\bibinfo {year} {2011})}\BibitemShut {NoStop}%
\bibitem [{\citenamefont {Brennan}\ \emph {et~al.}(2012)\citenamefont
  {Brennan}, \citenamefont {Cheong},\ and\ \citenamefont
  {Levchenko}}]{brennan2012information}%
  \BibitemOpen
  \bibfield  {author} {\bibinfo {author} {\bibfnamefont {M.~D.}\ \bibnamefont
  {Brennan}}, \bibinfo {author} {\bibfnamefont {R.}~\bibnamefont {Cheong}}, \
  and\ \bibinfo {author} {\bibfnamefont {A.}~\bibnamefont {Levchenko}},\
  }\href@noop {} {\bibfield  {journal} {\bibinfo  {journal} {Science}\ }\textbf
  {\bibinfo {volume} {338}},\ \bibinfo {pages} {334} (\bibinfo {year}
  {2012})}\BibitemShut {NoStop}%
\bibitem [{\citenamefont {Ladbury}\ and\ \citenamefont
  {Arold}(2012)}]{ladbury2012noise}%
  \BibitemOpen
  \bibfield  {author} {\bibinfo {author} {\bibfnamefont {J.~E.}\ \bibnamefont
  {Ladbury}}\ and\ \bibinfo {author} {\bibfnamefont {S.~T.}\ \bibnamefont
  {Arold}},\ }\href@noop {} {\bibfield  {journal} {\bibinfo  {journal} {Trends
  in biochemical sciences}\ }\textbf {\bibinfo {volume} {37}},\ \bibinfo
  {pages} {173} (\bibinfo {year} {2012})}\BibitemShut {NoStop}%
\bibitem [{\citenamefont {Voliotis}\ \emph {et~al.}(2014)\citenamefont
  {Voliotis}, \citenamefont {Perrett}, \citenamefont {McWilliams},
  \citenamefont {McArdle},\ and\ \citenamefont
  {Bowsher}}]{voliotis2014information}%
  \BibitemOpen
  \bibfield  {author} {\bibinfo {author} {\bibfnamefont {M.}~\bibnamefont
  {Voliotis}}, \bibinfo {author} {\bibfnamefont {R.~M.}\ \bibnamefont
  {Perrett}}, \bibinfo {author} {\bibfnamefont {C.}~\bibnamefont {McWilliams}},
  \bibinfo {author} {\bibfnamefont {C.~A.}\ \bibnamefont {McArdle}}, \ and\
  \bibinfo {author} {\bibfnamefont {C.~G.}\ \bibnamefont {Bowsher}},\
  }\href@noop {} {\bibfield  {journal} {\bibinfo  {journal} {Proceedings of the
  National Academy of Sciences}\ }\textbf {\bibinfo {volume} {111}},\ \bibinfo
  {pages} {E326} (\bibinfo {year} {2014})}\BibitemShut {NoStop}%
\bibitem [{\citenamefont {Chung}\ \emph {et~al.}(2010)\citenamefont {Chung},
  \citenamefont {Akita}, \citenamefont {Vandlen}, \citenamefont {Toomre},
  \citenamefont {Schlessinger},\ and\ \citenamefont
  {Mellman}}]{chung2010spatial}%
  \BibitemOpen
  \bibfield  {author} {\bibinfo {author} {\bibfnamefont {I.}~\bibnamefont
  {Chung}}, \bibinfo {author} {\bibfnamefont {R.}~\bibnamefont {Akita}},
  \bibinfo {author} {\bibfnamefont {R.}~\bibnamefont {Vandlen}}, \bibinfo
  {author} {\bibfnamefont {D.}~\bibnamefont {Toomre}}, \bibinfo {author}
  {\bibfnamefont {J.}~\bibnamefont {Schlessinger}}, \ and\ \bibinfo {author}
  {\bibfnamefont {I.}~\bibnamefont {Mellman}},\ }\href@noop {} {\bibfield
  {journal} {\bibinfo  {journal} {Nature}\ }\textbf {\bibinfo {volume} {464}},\
  \bibinfo {pages} {783} (\bibinfo {year} {2010})}\BibitemShut {NoStop}%
\bibitem [{\citenamefont {Schlessinger}(2000)}]{schlessinger2000cell}%
  \BibitemOpen
  \bibfield  {author} {\bibinfo {author} {\bibfnamefont {J.}~\bibnamefont
  {Schlessinger}},\ }\href@noop {} {\bibfield  {journal} {\bibinfo  {journal}
  {Cell}\ }\textbf {\bibinfo {volume} {103}},\ \bibinfo {pages} {211} (\bibinfo
  {year} {2000})}\BibitemShut {NoStop}%
\bibitem [{\citenamefont {Hill}(1998)}]{hill1998receptor}%
  \BibitemOpen
  \bibfield  {author} {\bibinfo {author} {\bibfnamefont {S.~M.}\ \bibnamefont
  {Hill}},\ }\href@noop {} {\bibfield  {journal} {\bibinfo  {journal} {The
  Anatomical Record: An Official Publication of the American Association of
  Anatomists}\ }\textbf {\bibinfo {volume} {253}},\ \bibinfo {pages} {42}
  (\bibinfo {year} {1998})}\BibitemShut {NoStop}%
\bibitem [{\citenamefont {Schwartz}\ and\ \citenamefont
  {Ginsberg}(2002)}]{schwartz2002networks}%
  \BibitemOpen
  \bibfield  {author} {\bibinfo {author} {\bibfnamefont {M.~A.}\ \bibnamefont
  {Schwartz}}\ and\ \bibinfo {author} {\bibfnamefont {M.~H.}\ \bibnamefont
  {Ginsberg}},\ }\href@noop {} {\bibfield  {journal} {\bibinfo  {journal}
  {Nature cell biology}\ }\textbf {\bibinfo {volume} {4}},\ \bibinfo {pages}
  {E65} (\bibinfo {year} {2002})}\BibitemShut {NoStop}%
\bibitem [{\citenamefont {Hunter}(2007)}]{hunter2007age}%
  \BibitemOpen
  \bibfield  {author} {\bibinfo {author} {\bibfnamefont {T.}~\bibnamefont
  {Hunter}},\ }\href@noop {} {\bibfield  {journal} {\bibinfo  {journal}
  {Molecular cell}\ }\textbf {\bibinfo {volume} {28}},\ \bibinfo {pages} {730}
  (\bibinfo {year} {2007})}\BibitemShut {NoStop}%
\bibitem [{\citenamefont {Kontogeorgaki}\ \emph {et~al.}(2017)\citenamefont
  {Kontogeorgaki}, \citenamefont {S{\'a}nchez-Garc{\'\i}a}, \citenamefont
  {Ewing}, \citenamefont {Zygalakis},\ and\ \citenamefont
  {MacArthur}}]{kontogeorgaki2017noise}%
  \BibitemOpen
  \bibfield  {author} {\bibinfo {author} {\bibfnamefont {S.}~\bibnamefont
  {Kontogeorgaki}}, \bibinfo {author} {\bibfnamefont {R.~J.}\ \bibnamefont
  {S{\'a}nchez-Garc{\'\i}a}}, \bibinfo {author} {\bibfnamefont {R.~M.}\
  \bibnamefont {Ewing}}, \bibinfo {author} {\bibfnamefont {K.~C.}\ \bibnamefont
  {Zygalakis}}, \ and\ \bibinfo {author} {\bibfnamefont {B.~D.}\ \bibnamefont
  {MacArthur}},\ }\href@noop {} {\bibfield  {journal} {\bibinfo  {journal}
  {Scientific Reports}\ }\textbf {\bibinfo {volume} {7}},\ \bibinfo {pages}
  {532} (\bibinfo {year} {2017})}\BibitemShut {NoStop}%
\bibitem [{\citenamefont {Ladbury}\ and\ \citenamefont
  {Arold}(2000)}]{ladbury2000searching}%
  \BibitemOpen
  \bibfield  {author} {\bibinfo {author} {\bibfnamefont {J.~E.}\ \bibnamefont
  {Ladbury}}\ and\ \bibinfo {author} {\bibfnamefont {S.}~\bibnamefont
  {Arold}},\ }\href@noop {} {\bibfield  {journal} {\bibinfo  {journal}
  {Chemistry \& biology}\ }\textbf {\bibinfo {volume} {7}},\ \bibinfo {pages}
  {R3} (\bibinfo {year} {2000})}\BibitemShut {NoStop}%
\bibitem [{\citenamefont {Ladbury}\ and\ \citenamefont
  {Arold}(2011)}]{ladbury2011energetics}%
  \BibitemOpen
  \bibfield  {author} {\bibinfo {author} {\bibfnamefont {J.~E.}\ \bibnamefont
  {Ladbury}}\ and\ \bibinfo {author} {\bibfnamefont {S.~T.}\ \bibnamefont
  {Arold}},\ }in\ \href@noop {} {\emph {\bibinfo {booktitle} {Methods in
  enzymology}}},\ Vol.\ \bibinfo {volume} {488}\ (\bibinfo  {publisher}
  {Elsevier},\ \bibinfo {year} {2011})\ pp.\ \bibinfo {pages}
  {147--183}\BibitemShut {NoStop}%
\bibitem [{\citenamefont {Levy}\ \emph {et~al.}(2010)\citenamefont {Levy},
  \citenamefont {Landry},\ and\ \citenamefont {Michnick}}]{levy2010signaling}%
  \BibitemOpen
  \bibfield  {author} {\bibinfo {author} {\bibfnamefont {E.~D.}\ \bibnamefont
  {Levy}}, \bibinfo {author} {\bibfnamefont {C.~R.}\ \bibnamefont {Landry}}, \
  and\ \bibinfo {author} {\bibfnamefont {S.~W.}\ \bibnamefont {Michnick}},\
  }\href@noop {} {\bibfield  {journal} {\bibinfo  {journal} {Science}\ }\textbf
  {\bibinfo {volume} {328}},\ \bibinfo {pages} {983} (\bibinfo {year}
  {2010})}\BibitemShut {NoStop}%
\bibitem [{\citenamefont {Breitkreutz}\ \emph {et~al.}(2010)\citenamefont
  {Breitkreutz}, \citenamefont {Choi}, \citenamefont {Sharom}, \citenamefont
  {Boucher}, \citenamefont {Neduva}, \citenamefont {Larsen}, \citenamefont
  {Lin}, \citenamefont {Breitkreutz}, \citenamefont {Stark}, \citenamefont
  {Liu} \emph {et~al.}}]{breitkreutz2010global}%
  \BibitemOpen
  \bibfield  {author} {\bibinfo {author} {\bibfnamefont {A.}~\bibnamefont
  {Breitkreutz}}, \bibinfo {author} {\bibfnamefont {H.}~\bibnamefont {Choi}},
  \bibinfo {author} {\bibfnamefont {J.~R.}\ \bibnamefont {Sharom}}, \bibinfo
  {author} {\bibfnamefont {L.}~\bibnamefont {Boucher}}, \bibinfo {author}
  {\bibfnamefont {V.}~\bibnamefont {Neduva}}, \bibinfo {author} {\bibfnamefont
  {B.}~\bibnamefont {Larsen}}, \bibinfo {author} {\bibfnamefont {Z.-Y.}\
  \bibnamefont {Lin}}, \bibinfo {author} {\bibfnamefont {B.-J.}\ \bibnamefont
  {Breitkreutz}}, \bibinfo {author} {\bibfnamefont {C.}~\bibnamefont {Stark}},
  \bibinfo {author} {\bibfnamefont {G.}~\bibnamefont {Liu}},  \emph {et~al.},\
  }\href@noop {} {\bibfield  {journal} {\bibinfo  {journal} {Science}\ }\textbf
  {\bibinfo {volume} {328}},\ \bibinfo {pages} {1043} (\bibinfo {year}
  {2010})}\BibitemShut {NoStop}%
\bibitem [{\citenamefont {Kinney}\ and\ \citenamefont
  {Atwal}(2014)}]{kinney2014equitability}%
  \BibitemOpen
  \bibfield  {author} {\bibinfo {author} {\bibfnamefont {J.~B.}\ \bibnamefont
  {Kinney}}\ and\ \bibinfo {author} {\bibfnamefont {G.~S.}\ \bibnamefont
  {Atwal}},\ }\href@noop {} {\bibfield  {journal} {\bibinfo  {journal}
  {Proceedings of the National Academy of Sciences}\ ,\ \bibinfo {pages}
  {201309933}} (\bibinfo {year} {2014})}\BibitemShut {NoStop}%
\bibitem [{\citenamefont {Tareen}\ \emph {et~al.}(2018)\citenamefont {Tareen},
  \citenamefont {Wingreen},\ and\ \citenamefont
  {Mukhopadhyay}}]{tareen2018modeling}%
  \BibitemOpen
  \bibfield  {author} {\bibinfo {author} {\bibfnamefont {A.}~\bibnamefont
  {Tareen}}, \bibinfo {author} {\bibfnamefont {N.~S.}\ \bibnamefont
  {Wingreen}}, \ and\ \bibinfo {author} {\bibfnamefont {R.}~\bibnamefont
  {Mukhopadhyay}},\ }\href@noop {} {\bibfield  {journal} {\bibinfo  {journal}
  {Physical Review E}\ }\textbf {\bibinfo {volume} {97}},\ \bibinfo {pages}
  {020402} (\bibinfo {year} {2018})}\BibitemShut {NoStop}%
\bibitem [{\citenamefont {Mok}\ \emph {et~al.}(2010)\citenamefont {Mok},
  \citenamefont {Kim}, \citenamefont {Lam}, \citenamefont {Piccirillo},
  \citenamefont {Zhou}, \citenamefont {Jeschke}, \citenamefont {Sheridan},
  \citenamefont {Parker}, \citenamefont {Desai}, \citenamefont {Jwa} \emph
  {et~al.}}]{mok2010deciphering}%
  \BibitemOpen
  \bibfield  {author} {\bibinfo {author} {\bibfnamefont {J.}~\bibnamefont
  {Mok}}, \bibinfo {author} {\bibfnamefont {P.~M.}\ \bibnamefont {Kim}},
  \bibinfo {author} {\bibfnamefont {H.~Y.}\ \bibnamefont {Lam}}, \bibinfo
  {author} {\bibfnamefont {S.}~\bibnamefont {Piccirillo}}, \bibinfo {author}
  {\bibfnamefont {X.}~\bibnamefont {Zhou}}, \bibinfo {author} {\bibfnamefont
  {G.~R.}\ \bibnamefont {Jeschke}}, \bibinfo {author} {\bibfnamefont {D.~L.}\
  \bibnamefont {Sheridan}}, \bibinfo {author} {\bibfnamefont {S.~A.}\
  \bibnamefont {Parker}}, \bibinfo {author} {\bibfnamefont {V.}~\bibnamefont
  {Desai}}, \bibinfo {author} {\bibfnamefont {M.}~\bibnamefont {Jwa}},  \emph
  {et~al.},\ }\href@noop {} {\bibfield  {journal} {\bibinfo  {journal} {Sci.
  Signal.}\ }\textbf {\bibinfo {volume} {3}},\ \bibinfo {pages} {ra12}
  (\bibinfo {year} {2010})}\BibitemShut {NoStop}%
\bibitem [{\citenamefont {Fischbach}\ \emph {et~al.}(2013)\citenamefont
  {Fischbach}, \citenamefont {Bluestone},\ and\ \citenamefont
  {Lim}}]{fischbach2013cell}%
  \BibitemOpen
  \bibfield  {author} {\bibinfo {author} {\bibfnamefont {M.~A.}\ \bibnamefont
  {Fischbach}}, \bibinfo {author} {\bibfnamefont {J.~A.}\ \bibnamefont
  {Bluestone}}, \ and\ \bibinfo {author} {\bibfnamefont {W.~A.}\ \bibnamefont
  {Lim}},\ }\href@noop {} {\bibfield  {journal} {\bibinfo  {journal} {Science
  translational medicine}\ }\textbf {\bibinfo {volume} {5}},\ \bibinfo {pages}
  {179ps7} (\bibinfo {year} {2013})}\BibitemShut {NoStop}%
\bibitem [{\citenamefont {Bashor}\ and\ \citenamefont
  {Collins}(2018)}]{bashor2018understanding}%
  \BibitemOpen
  \bibfield  {author} {\bibinfo {author} {\bibfnamefont {C.~J.}\ \bibnamefont
  {Bashor}}\ and\ \bibinfo {author} {\bibfnamefont {J.~J.}\ \bibnamefont
  {Collins}},\ }\href@noop {} {\bibfield  {journal} {\bibinfo  {journal}
  {Annual review of biophysics}\ }\textbf {\bibinfo {volume} {47}},\ \bibinfo
  {pages} {399} (\bibinfo {year} {2018})}\BibitemShut {NoStop}%
\end{thebibliography}%

\bibliographystyle{apsrev4-1}


\widetext
\newpage
\begin{center}
	\textbf{\large Supplemental Information for ``The strength of protein-protein interactions controls the information capacity and dynamical response of signaling network"}
\end{center}

\author{Ching-Hao Wang}
\email{chinghao@bu.edu}
\affiliation{Department of Physics and Biological Design Center, Boston University, Boston, MA 02215, USA}

\author{Caleb J. Bashor}
\email{caleb.bashor@rice.edu}
\affiliation{Department of Bioengineering, Rice University, Huston, TX 77030, USA}

\author{Pankaj Mehta}
\email{pankajm@bu.edu}
\affiliation{Department of Physics and Biological Design Center, Boston University, Boston, MA 02215, USA}

\date{\today}

\maketitle
\setcounter{section}{0}
\setcounter{equation}{0}
\setcounter{figure}{0}
\setcounter{table}{0}
\setcounter{page}{1}

\makeatletter
\renewcommand{\theequation}{S\arabic{equation}}
\renewcommand{\thefigure}{S\arabic{figure}}


	\section{Signal-to-noise ratio (SNR) of signaling circuit}
	\label{sec:SI-SNR}
	Consider a biochemical pathway that involves the relay of a (possibly continuous) signal $c$ to the intracellular kinase $X$ which intern activates an output $Y$. We assume that these proteins are catalytically active only when they undergo a post-translational modification (PTM). We represent the PTM-state of the proteins  as binary random variables that take value $1$ when catalytically active and $0$ otherwise. Pictorially, this pathway can be summarized as the following channel: $c\rightarrow X\rightarrow Y$. Note that in the appendix, we first set $\beta=1$ to simplify notation in the calculation and put it back in at the end using dimensional analysis.
	
\begin{figure}
	\begin{center}
		\includegraphics[width=0.5 \textwidth]{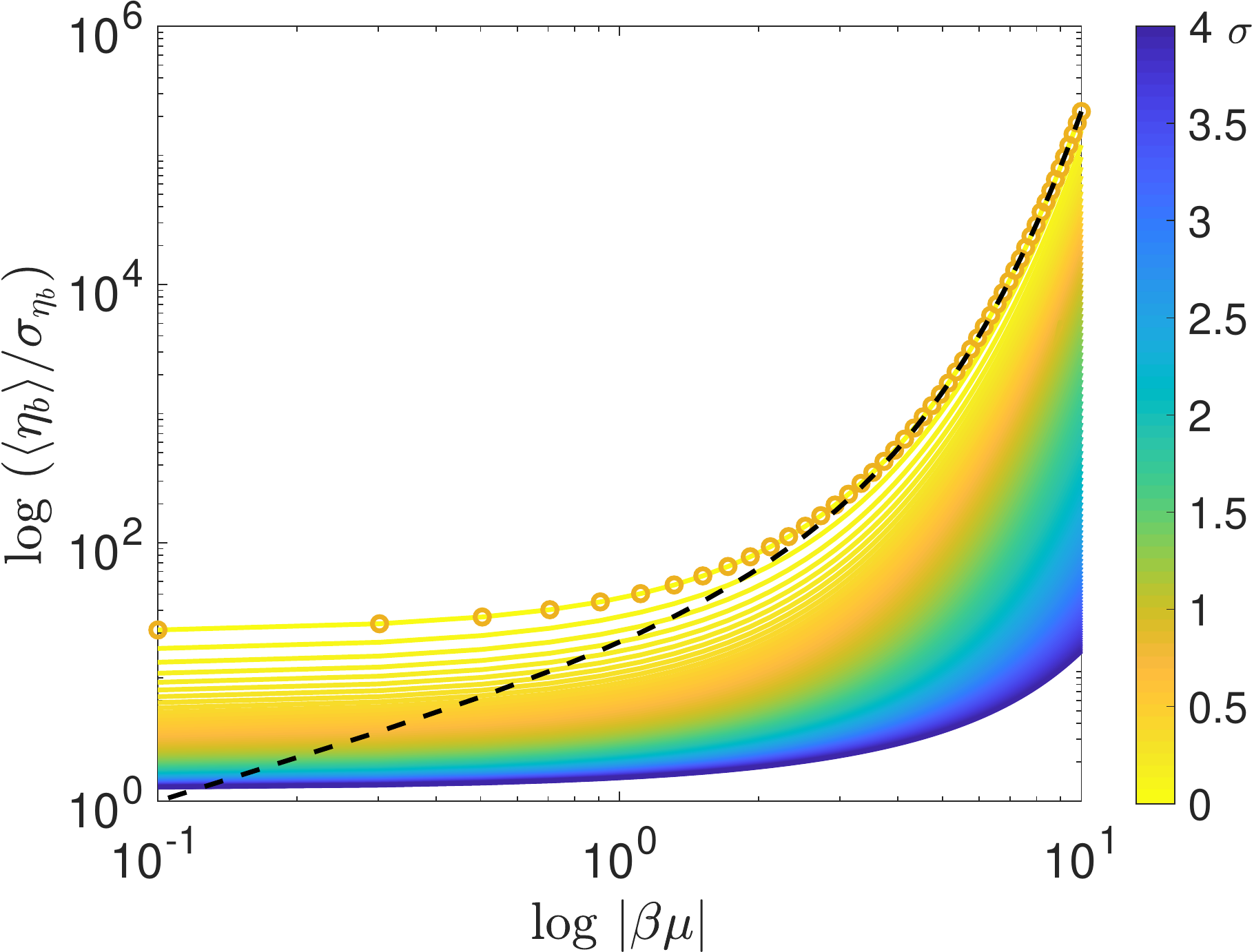}
	\caption{Inverse coefficient of variation (CV$^{-1}$) of $Q(c)$ assuming negligible $\eta_W$ is shown for different value of  $\mu\equiv \langle \theta_{i,j}\rangle$ and $\sigma^2 \equiv \text{Var}(\theta_{i,j})$, where $\theta_{i,j}$ is the binding affinity of $K_i$ to $K_j$ and the average $\langle\cdot\rangle$ is taken over different parameter realizations $\theta_{i,j}$. Note that $\mu <0$ by definition. Different colors indicate different $\sigma$ whose values are encoded in the color bar. Colored curves are theoretical predictions Eq.\eqref{eq:SI:CVnum}\eqref{eq:SI:CVden}, open circles are calculated via sampling network parameters as described in \emph{Materials and Methods}. Black dashed curve is the analytical result in the tight-binding limit whose expression is given by Eq.\eqref{eq:eta_1stmo}\eqref{eq:eta_var} with $\sigma$ corresponding to that of the yellow open circles ($\sigma = 0.1$). Note the logarithmic scale.}
	\label{fig:SI-theta-var}
\end{center}
	\end{figure}

	To calculate the signal-to-noise ratio (SNR),we need the probability of the output given the input, $Q(c)\equiv \P(Y=1|c)$. Specifically,
	\bea
	\P(Y=1|c) &=& \sum_{X\in\{0,1\}}\P(Y=1|X)\P(X|c)\\
	&=& \underbrace{\left(\frac{e^{-\theta_x}}{1+e^{-\theta_x}}\right)}_{\eta_b}q(c) + \underbrace{\left(\frac{e^{-W}}{1+e^{-W}}\right)}_{\eta_W}\left(1-q(c)\right),
	\eea
	where $q(c)\equiv \P(X=1|c)$ is the probability that kinase $X$ is phosphorylated in the presence of a ligand at concentration $c$. The functional form of $q(c)$ is not relevant so we simply assume that it is a monotonically increasing function of $c$ and attains 1(0) when $c=1(0)$. Note that when $q(c)=1$ (i.e. full input signal), $Q(c)$ is purely dictated by bindings of phosphorylated kinase to its substrate (i.e. $\theta_x$, see Fig.~\ref{fig:network-model}D), whereas when $q(c)=0$ (i.e. no input signal), the contribution is solely from those involving unphosphorylated kinase (i.e. $W$, see Fig.~\ref{fig:network-model}D). Therefore, we define the signal-to-noise ratio (SNR) formally as:
	\be
	\text{SNR}\equiv \frac{\langle Q(c=1)\rangle}{\langle Q(c=0)\rangle} =\frac{\langle \eta_b \rangle}{\langle \eta_W \rangle},
	\ee
	where $\langle\cdot \rangle$ denotes the average with respect the distribution of BAs. To simplify, we assume that $W$ is a constant that sets the time scale of non-specific bindings and that the specific BA $\theta_x\sim \mathcal{N}(\mu,\sigma^2)$ is drawn from a Gaussian distribution with mean $\mu \ll -1$ (i.e. tight-binding) and variance $\sigma^2$. Since $\theta_x$ is normally distributed, $e^{\theta_x}$ follows log-normal distribution. In this tight-binding approximation, 
	\be
	\eta_b=\frac{1}{1+e^{\theta_x}}\approx 1-e^{\theta_x}\equiv 1-Z,
	\ee
	where $Z\equiv e^{\theta_x}\sim\log\mathcal{N}(\mu,\sigma^2)$. From this, one can calculate its first two moments:
	\bea
	\label{eq:eta_1stmo}
	\langle\eta_b\rangle=\mathbb{E}[1-Z] &=&\int_0^\infty dz\,(1-z)\frac{1}{z\sigma\sqrt{2\pi}}\exp\left[-\frac{\left(\log z - \mu\right)^2}{2\sigma^2}\right]\nonumber\\
	&=& 1-\exp\left(\mu+\frac{\sigma^2}{2}\right)
	\eea
	and
	\bea
	\label{eq:eta_2ndmo}
	\langle\eta_b^2\rangle=\mathbb{E}[(1-Z)^2] &=&\int_0^\infty dz\, \frac{1-2z+z^2}{\sigma\sqrt{2\pi}}\exp\left[-\frac{\left(\log z - \mu\right)^2}{2\sigma^2}\right]\nonumber\\
	&=& 1-2\exp\left(\mu+\frac{\sigma^2}{2}\right) + \exp [2 \left(\mu+\sigma^2\right)].
	\eea
	From this one can also derive its variance
	\bea
	\label{eq:eta_var}
	\text{Var}(\eta_b) &=& \langle\eta_b^2\rangle-\langle\eta_b\rangle^2\nonumber\\
	&=& e^{2(\mu+\sigma^2)}(1-e^{-\sigma^2}).
	\eea
	These quantities can be used to analyze the effect of heterogeneity in $\theta_X$ on $Q(c)$ which we summarized in Fig.~\ref{fig:SI-theta-var}. Finally, after putting the energy unit $\beta^{-1}=k_BT$ back in and noting that $e^{-\beta W}\ll 1$ so that $\eta_W\approx e^{-\beta W}$, the signal-to-noise ratio is simply
	\be
	\text{SNR}=\frac{\langle \eta_b\rangle}{\langle \eta_W\rangle} = e^{\beta W}\left[1-e^{\beta\left(\mu+\frac{\sigma^2}{2}\right)}\right].
	\ee
	Note that one can still calculate the SNR without assuming tight-binding, except in this case there's no closed form solution. Follow the same procedure while retaining $\eta_b =e^{-\theta_x}/(1+e^{-\theta_x})$ and performing change-of-variable, one ended up with the following integrals:
	\bea\label{eq:SI:CVnum}
	\langle \eta_b\rangle &=&\frac{1}{\sigma \sqrt{2\pi}}\int_0^\infty d\lambda\, \frac{1}{1+\lambda}\exp\left[-\frac{\left(\log \lambda + \mu\right)^2}{2\sigma^2}\right]\\\label{eq:SI:CVden}
	\langle \eta_b^2\rangle &=&\frac{1}{\sigma \sqrt{2\pi}}\int_0^\infty d\lambda\, \frac{\lambda}{(1+\lambda)^2}\exp\left[-\frac{\left(\log \lambda + \mu\right)^2}{2\sigma^2}\right].
	\eea
	One can further apply Laplace method by assuming $M\equiv 1/(2\sigma^2)\gg 0$ (i.e. $\sigma^2\rightarrow 0$, zero temperature limit) to get
	\bea
	\langle \eta_b\rangle_{M\gg 1} &\approx& \frac{e^{-\mu}}{1+e^{-\mu}} \\
	\langle \eta_b^2\rangle_{M\gg 1} &\approx& \left(\frac{e^{-\mu}}{1+e^{-\mu}}\right)^2,
	\eea
	implying that Var$(\eta_b)\sim 0$ and $\langle \eta_b\rangle$ is simply the mean-field value. In this case, the SNR with the energy unit in place reads 
	\be
	\text{SNR}_{M\gg 1} \approx \frac{e^{-\beta(\mu-W)}}{1+e^{-\beta\mu}}.
	\ee
	
	\section{Deriving the information capacity}
	\label{sec:SImodel}
	In this section, we derive the mutual information transduced across a linear signaling network based on phosphorylation cascade. Note that linearity here refers to the network topology not that of the transfer function relating phosphorylation reaction downstream. Concretely, we consider a $n$-phosphorylation kinase cascade represented by a Bayesian network of the form: $x_\text{in}\rightarrow x_1\rightarrow\cdots\rightarrow x_n \rightarrow x_\text{out}$, where $x_i, x_\text{in}, x_\text{out}\in\{0,1\}$ as defined in the main text.  For brevity, we denote $x_\text{out}\equiv x_{n+1}$ and $x_\text{in}\equiv x_0$. Due to the Markovian nature of this network, the joint distribution of kinase states can be factorized as
	\be\label{eq:joint}
	\P(\mbf{x}) := \P( x_0,\cdots, x_{n+1} )=\left(\prod_{i=0}^{n}\P(x_{i+1}|x_i)\right) \P(x_0),
	\ee
	where the conditionals are given by
	\be\label{eq:boltzmann}
	\P(x_{i+1}=1|x_i)=\frac{x_i e^{-\theta_{i,i+1}}}{1+x_i e^{-\theta_{i,i+1}}}.
	\ee
	Note that we denote the \emph{relative binding affinity} of $i$ to $i+1$ as $\theta_{i, i+1}$ (c.f. Eq. \eqref{eq:theta-kinetics}) . From now on, every energetic parameters are measured in units of $k_BT$. To simplify notation, we represent the conditional probability by a transfer matrix defined as:
	\be\label{eq:M}
	\M_{i+1,i} =\begin{pmatrix}
		\P(x_{i+1}=1|x_i=1) & \P(x_{i+1}=1|x_i=0)\\
		\P(x_{i+1}=0|x_i=1) & \P(x_{i+1}=0|x_i=0)
	\end{pmatrix}
	\ee 
	To calculate the mutual information between $x_1$ and $x_n$, 
	\be\label{eq:MI}
	I(x_0; x_{n+1}) =\sum_{x_0}\sum_{x_{n+1}}\P(x_0)\P(x_{n+1}|x_0)\log_2\left[\frac{\P(x_{n+1}|x_0)}{\P(x_{n+1})}\right],
	\ee
	we need to get $\P(x_{n+1}|x_0)$ first. Using the matrix notation, we have
	\bea\label{eq:Pn1}
	\P(x_{n+1}|x_0)&=&\sum_{x_1}\cdots\sum_{x_{n}}\prod_{i=0}^{n}\P(x_{i+1}|x_i)\nonumber\\
	&=&\sum_{x_{n}}\P(x_{n+1}|x_{n})\P(x_{n}|x_{n-1})\cdots\sum_{x_2}\P(x_3|x_2)\sum_{x_1}\P(x_2|x_1)\P(x_1|x_0)\nonumber\\
	&=&\prod_{i=0}^{n} \M_{i+1,i}\equiv \mathbf{P}_{n+1,0}, 
	\eea
	from which we can derive the marginal probability of $x_{n+1}$:
	\be\label{eq:pn}
	\mbf{p}_{n+1}= \sum_{x_0}\P(x_{n+1}|x_0)\P(x_0) = \left(\prod_{i=0}^{n} \M_{i+1,i}\right) \mbf{p}_0 = \mbf{P}_{n+1,0}\mbf{p}_0,
	\ee
	where $\mbf{p}_{n+1}\equiv (\P(x_{n+1}=1), \P(x_{n+1}=0))^T$, and similarly for $\mbf{p}_0$. With this defined and for a given set of $\theta_{i,j}$, we can calculate mutual information Eq. \eqref{eq:MI} by a series of matrix multiplications. 
	
\begin{figure}
	\begin{center}
		\includegraphics[width=0.5 \textwidth]{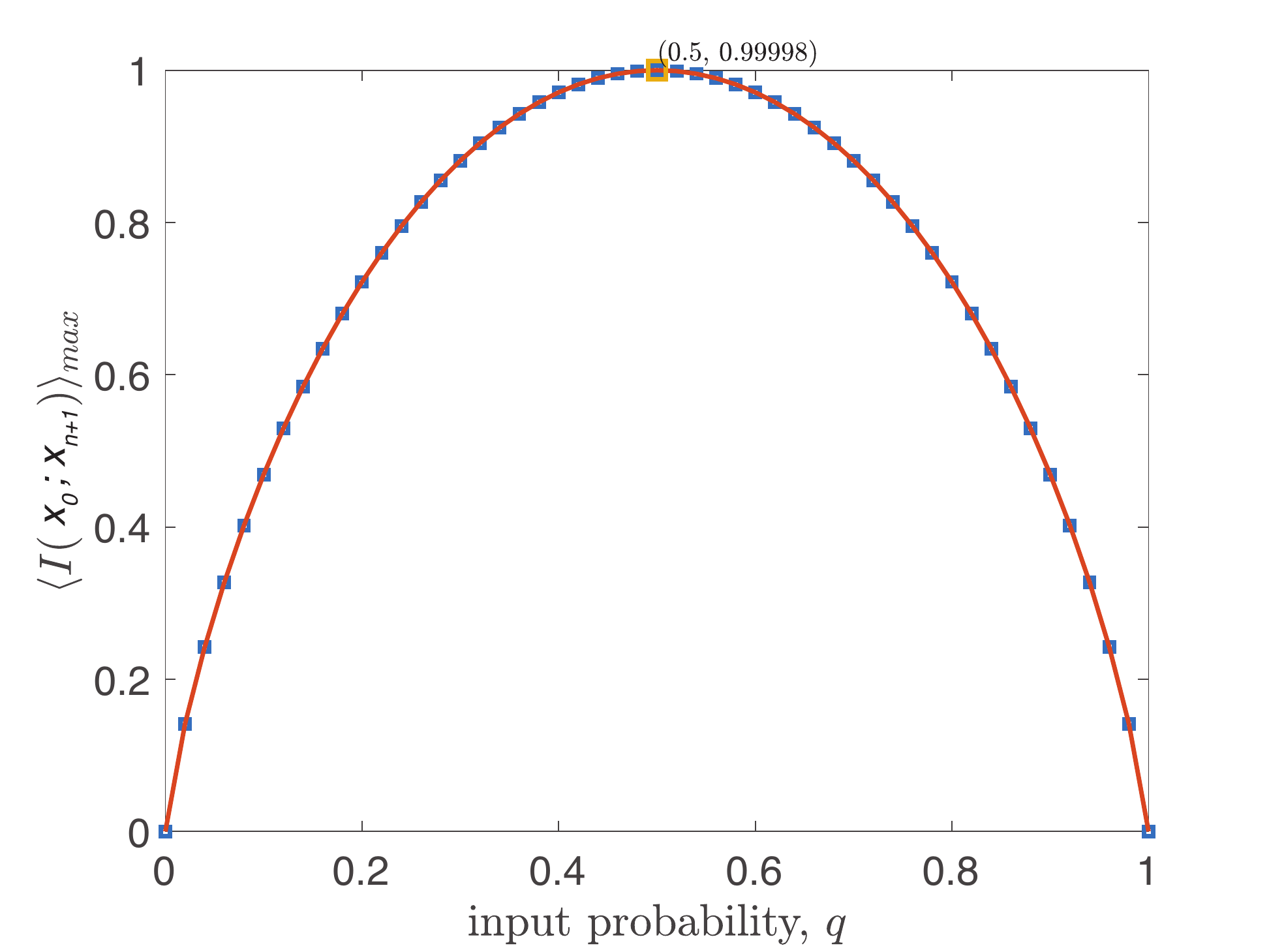}
	\caption{At tight-binding mutual information for linear network reduces to binary entropy function. Linear network of depth $n=8$ is used. Red squares are obtained by averaging the result over 100 different realizations of binding affinities using the methods detailed in this appendix. Dashed black curve is plotted using Eq. \eqref{eq:MI_lin_TB}. Parameters used are: $\beta\mu=-5, \sigma = 0.1$. The value and location of maximum mutual information obtained by averaging is indicated as $(q, I_\text{max})= (0.5, 0.99998)$.}
	\label{fig:SI-MI_lin_max}
\end{center}
\end{figure}

	Now consider several realizations of such signaling circuits with $\theta_{i,j}$ drawn from some distribution, say, Gaussian with mean $\mu$ and variance $\sigma^2$. In the tight-binding limit, $\mu \ll -1$ and the transfer matrix Eq. \eqref{eq:M} approximates the identity matrix $\mbf{I}$ due to Eq. \eqref{eq:boltzmann}. From this, one can easily show that mutual information averaged over different realizations is given by:
	\be\label{eq:MI_lin_TB}
	\langle I(x_0; x_{n+1})\rangle =-q \log_2 q -(1-q)\log_2(1-q) \equiv H_2(q),
	\ee
	where $q\equiv \P(x_1=1)$, and $H_2(q)$ is the entropy function of Bernoulli process with probability $q$ of one of the two values. Note that this calculation does not depend on the depth of the network (i.e. $n$), which implies as long as this approximation holds (i.e. tight-binding), mutual information is always peaked when input is least certain (i.e. $q=0.5$), see Figure~\ref{fig:SI-MI_lin_max}.

	\begin{figure}
	\begin{center}
		\includegraphics[width=0.8 \textwidth]{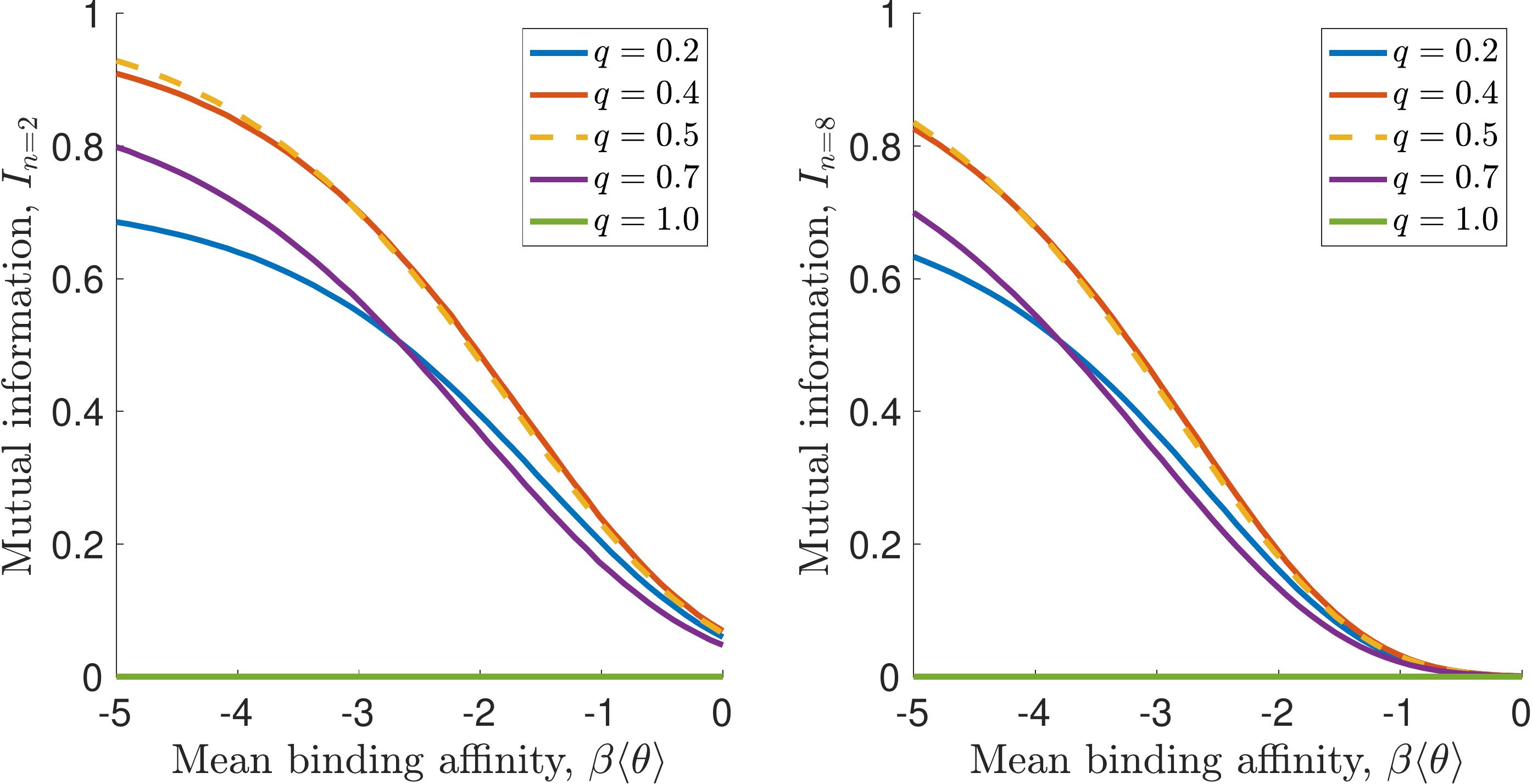}
	\caption{Mutual information as a function of mean binding affinity $\mu\equiv \langle\theta\rangle$ at different inputs. Note that maximum mutual information may not occur at $q=0.5$ (dashed orange curve) away from tight-binding (i.e. less negative $\beta\langle\theta\rangle$). In all panels, $\sigma = 0.1$. }
	\label{fig:SI-MI-v-mu}
\end{center}
\end{figure}

	\section{Optimal input to reach maximum mutual information}
	In this section, we derive the optimal input that gives maximum mutual information. To simplify notation, let's define 
	\be
	b_n=\prod_{i=1}^{n-1} f_i =\prod_{i=1}\left(\frac{e^{-\theta_{i,i+1}}}{1+e^{-\theta_{i,i+1}}}\right).
	\ee
	From this, one can re-write Eq. \eqref{eq:Pn1} and Eq. \eqref{eq:pn} as 
	\bea
	\mbf{P}_{n,1} = \begin{pmatrix}
		b_n & 0 \\
		1-b_n& 1
	\end{pmatrix}, \qquad
	\mbf{p}_n = \begin{pmatrix}
		q \,b_n  \\
		1-q \,b_n 
	\end{pmatrix}.
	\eea
	Plugging this back in to the definition of mutual information,Eq.  \eqref{eq:MI}, one gets,
	\be\label{eq:MI_linear_explicit}
	I(x_1;x_n)=-q b_n\log q + q (1-b_n)\log (1-b_n) - (1- q\, b_n )\log (1- q\, b_n )
	\ee
	After taking the derivate of Eq. \eqref{eq:MI_linear_explicit} with respect to $q$ and setting it to zero, one finds that the optimal input $q^\star$ that gives the maximum mutual information $I(x_1;x_n)$ is the solution to the following transcendental equation:
	\be\label{eq:qopt_linear}
	b_n\log\left(\frac{q^\star}{1-q^\star b_n }\right)=(1-b_n)\log (1-b_n)
	\ee
	
	\begin{figure}
	\begin{center}
		\includegraphics[width=0.8 \textwidth]{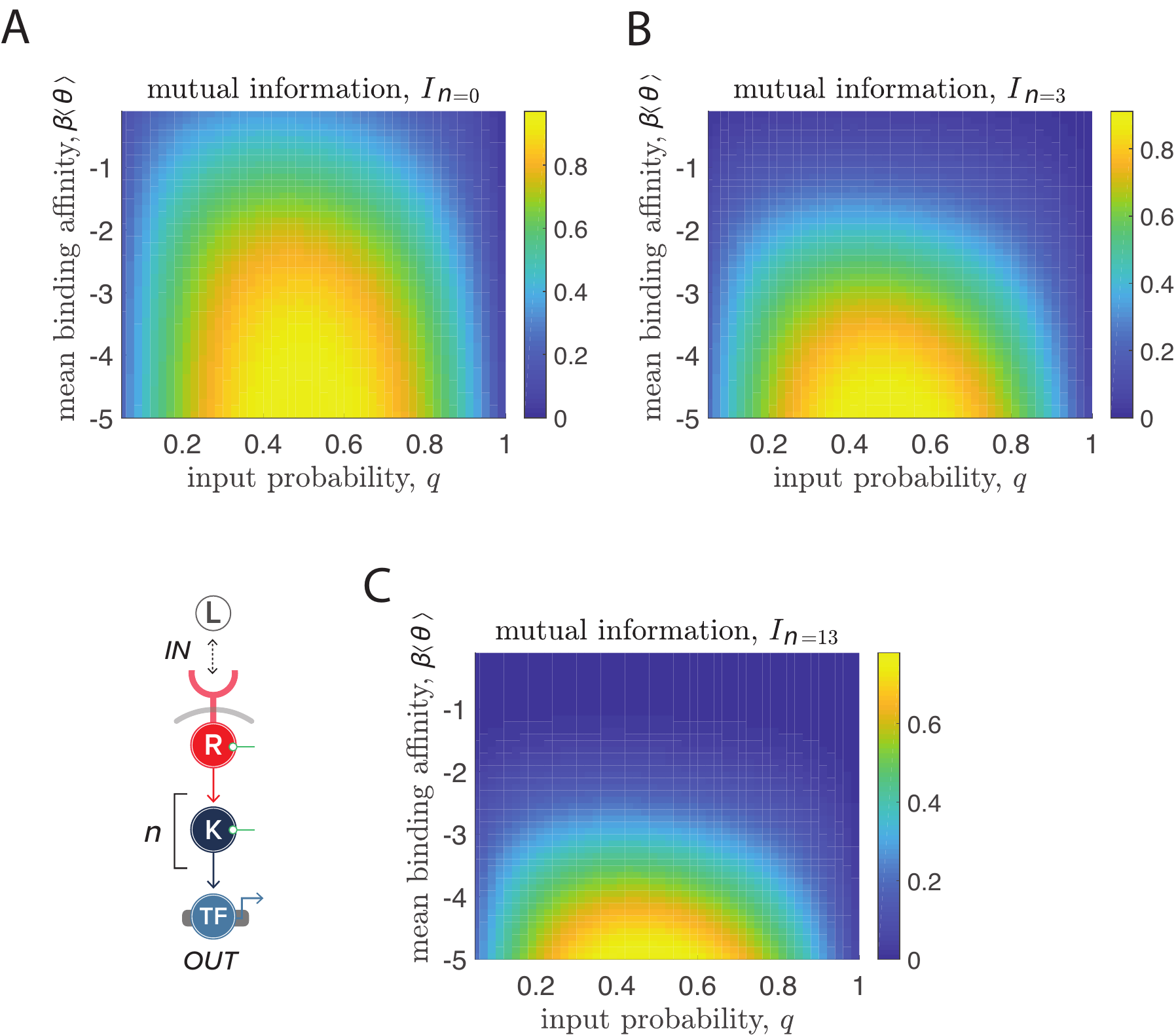}
	\caption{{\bf Tighter binding always increases information transmission for cascades of depths $n=0,3,13$}.  Here Var$(\theta_{i,j})\equiv \sigma^2=0.01$. All panels are generated by averaging 100 realizations of binding affinities using the scheme detailed in SI Section 2. }
	\label{fig:SI-linear_chain_pd_n}
\end{center}
\end{figure}

	\begin{figure}
	\begin{center}
		\includegraphics[width=0.5 \textwidth]{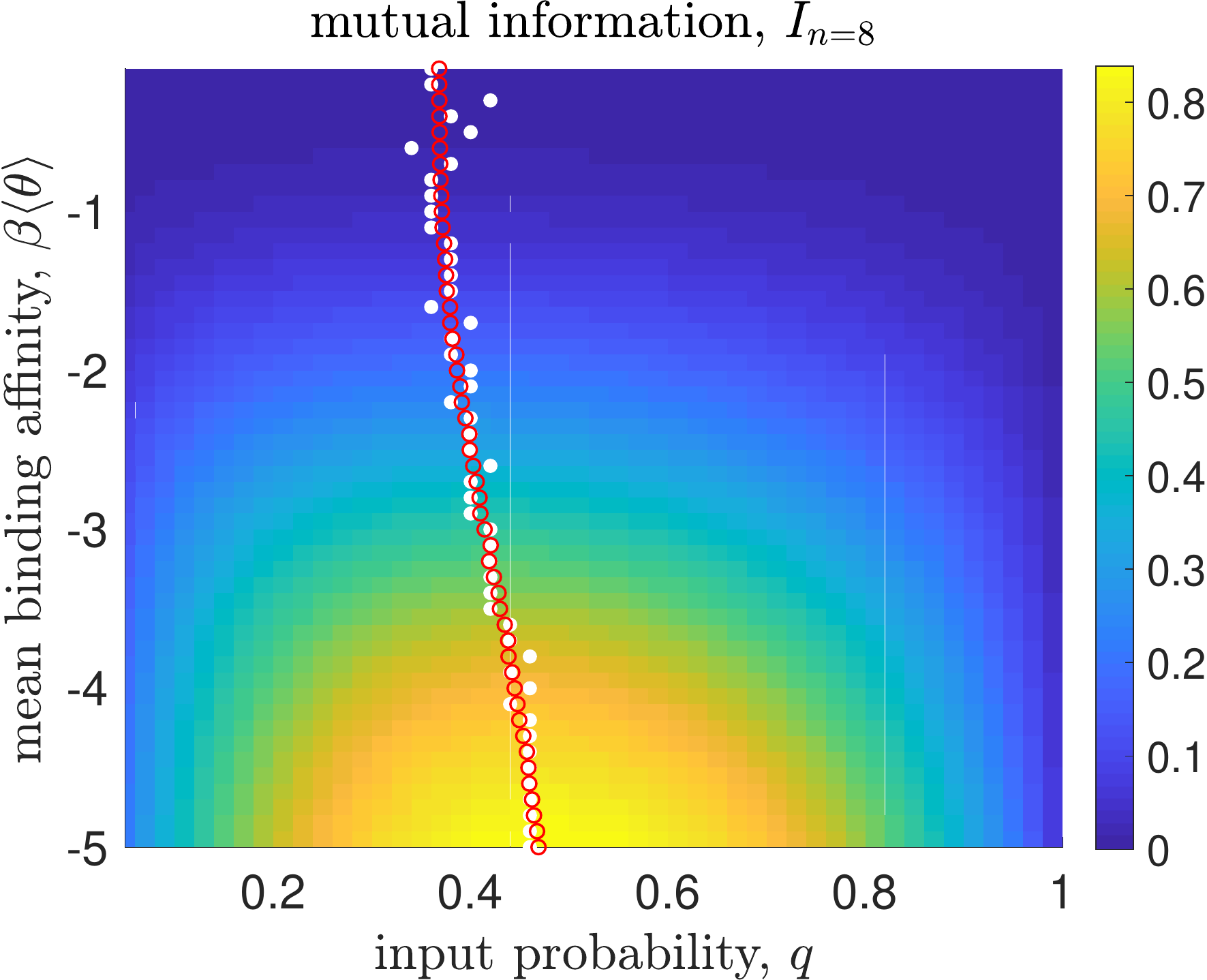}
	\caption{{\bf Location of maximum mutual information is not necessarily at input with most uncertainty ($q=0.5$)}. White filled circles are calculated by numerically searching for the input $q^\star$ on this color map that gives maximum mutual information. Red open circles are obtained through solving Eq. \eqref{eq:qopt_linear}. Here Var$(\theta_{i,j})\equiv \sigma^2=0.01$. The color map is generated by averaging 100 realizations of binding affinities using the scheme detailed in SI Section 2.}
	\label{fig:SI-linear_chain_pd_sol}
\end{center}
\end{figure}


	\section{Relating thermodynamics to a kinetic model of phosphorylation cascade}
	\begin{figure}
	\begin{center}
		\includegraphics[width=10cm]{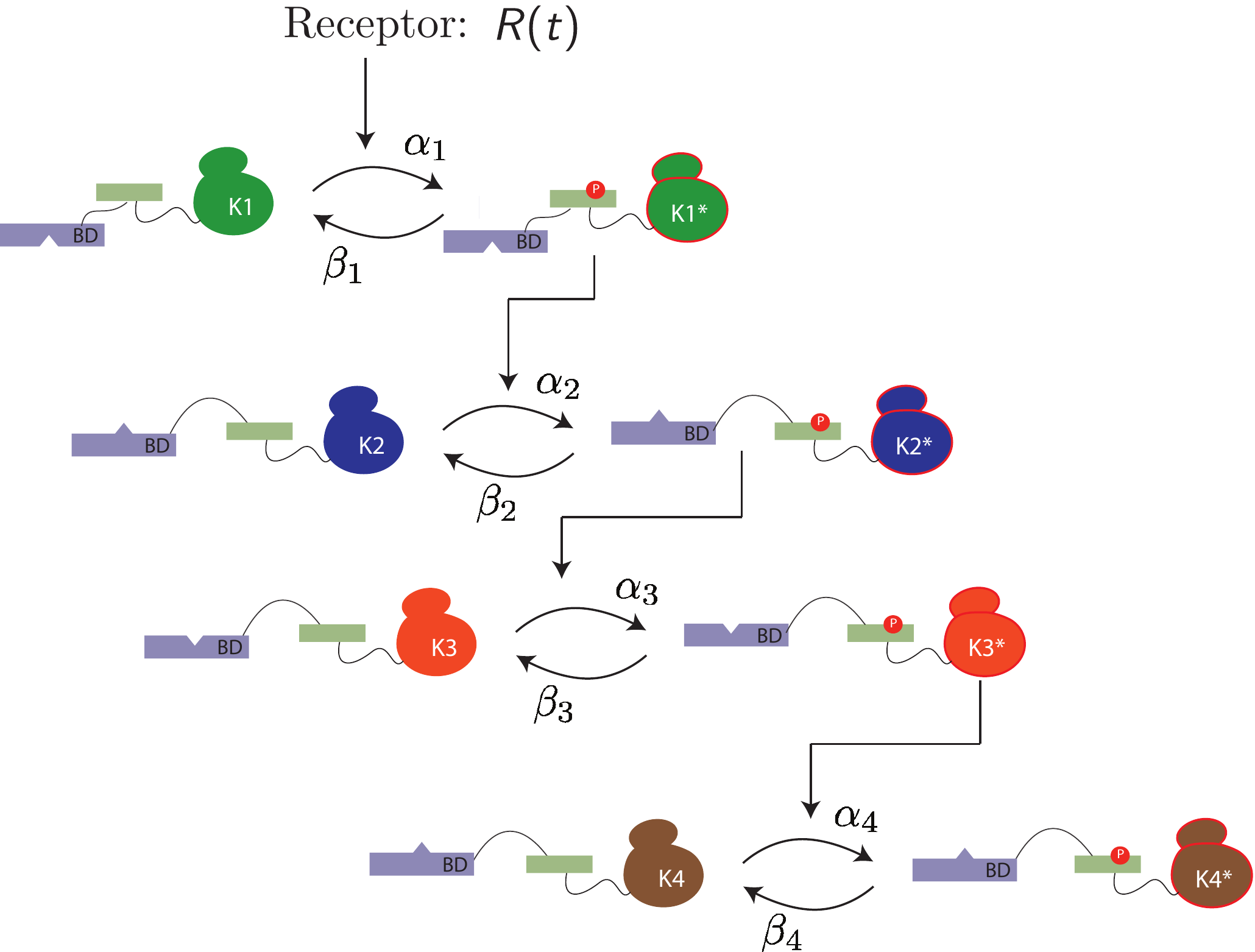}
	\caption{{\bf Signal network based on PK cascade. } The input of this signaling pathway is conceptualized as receptor kinase activation $R$ which could potentially be time-varying. The phosphorylation cascade depicted here is as described in the main text. Here we denote $\alpha_i$ ($\beta_i$) as the phosphorylation (dephosphorylation) rate of the cascade stage $K_i\rightarrow K_{i+1}$.  }
	\label{fig:SI-PKmodel}
\end{center}
\end{figure}

	Here we derive the Eq.\eqref{eq:MIdef} in the main text  (re-written as Eq. \eqref{eq:SI-SSprob} here) from chemical kinetics. Following Fig.~\ref{fig:SI-PKmodel}, let $X_i$ be the concentration of kinase $i$ in its active (i.e. phosphorylated) form and $\tilde{X}_i$ be that of its inactive (i.e. unphosphorylated) form. For each step $i$ of cascade except for $i=1$, the rate of phosphorylation is dependent on the concentration of active kinase $X_{i-1}$ and that of the inactive downstream  $\tilde{X}_i$. We describe the phosphorylation rate of kinase $i$ by  $\Phi_i^+=\tilde{\alpha}_i X_{i-1}\tilde{X}_i$. Assuming the phosphatase concentration is constant, we can write down the dephosphorylation rate as $\Phi_i^- = \beta_i X_i$. Here $\tilde{\alpha}_i,\,\beta_i$ are the kinetics rate constants of phosphorylation and dephosphorylation reactions, respectively. With this defined we can write down the kinetics equations for all kinases in the pathway (except for the first one) as: ($\forall i>1$)
	\bea
	\frac{dX_i}{dt}& =& \Phi_i^+-\Phi_i^-\\
	&=& \tilde{\alpha}_i X_{i-1}\tilde{X}_i - \beta X_i \\
	&=& \alpha_iX_{i-1}\left(1-\frac{X_i}{C_i}\right)-\beta_i X_i,
	\eea
	where $\alpha_i=\tilde{\alpha}_iC_i$ is the pseudo-first order rate constant and $C_i = X_i+\tilde{X}_i$ is the total concentration of kinase $i$. For the first kinase, its phosphorylation is stimulated by active receptors whose concentration is denoted as $R(t)$. In addition, it is dephosphorylated by phosphatase at rate $\beta_1$. Combining this we have
	\be
	\frac{dX_1}{dt} = \alpha_1R(t)\left(1-\frac{X_1}{C_1}\right)-\beta_1 X_1.
	\ee
	At steady-state, we have for $i\neq 1$
	\be\label{eq:SI-Xss}
	X_i^{SS}=\frac{C_i X_{i-1}^{SS}}{\gamma_i C_i + X_{i-1}^{SS}},
	\ee
	where $\gamma_i =\beta_i/\alpha_i$. Divide both sides by the total concentration of kinase $i$, $C_i$, one gets the steady-state activation probability of $i$:
	\be\label{eq:SI-SSprob-kinetics}
	p_i^{SS}=\frac{C_iX_{i-1}^{SS}(\gamma_i C_i)^{-1}}{1+ X_{i-1}^{SS}(\gamma_i C_i)^{-1}}.
	\ee
	Eq.\eqref{eq:SI-Xss} is related to the Michaelis-Menton equation $V_{\text{max}}S/(K_m+S)$ by recognizing 
	\bea
	X_{i-1}^{SS}&\rightarrow& S \\
	C_i&\rightarrow& V_{\text{max}}\\
	K_m &\rightarrow& C_i\gamma_i
	\eea
	
	Finally, the steady-state probability model presented in the main text,
	\bea\label{eq:SI-SSprob}
	\P(x_i=1|x_{i-1})&=&\frac{x_{i-1}e^{-\theta_{i-1,i}/(k_BT)}}{1+x_{i-1}e^{-\beta\theta_{i-1,i}/(k_BT)}},
	\eea
	can be interpreted under this kinetic framework by relating
	\be\label{eq:theta-kinetics}
	\theta_{i-1,i} =k_BT\ln \left(\frac{K_m}{X_{i-1}^{SS}}\right)= \Delta F -\tilde{\mu}_{i-1},
	\ee
	where $\Delta F = k_BT \ln K_m$ is the free energy difference between the bound and unbound state and $\tilde{\mu}_{i-1}= k_BT\ln  X_{i-1}^{SS}$ is the chemical potential of active kinase $i-1$.


	\section{Effects of network depth}
\begin{figure}
		\begin{center}
	\includegraphics[width=0.9 \textwidth]{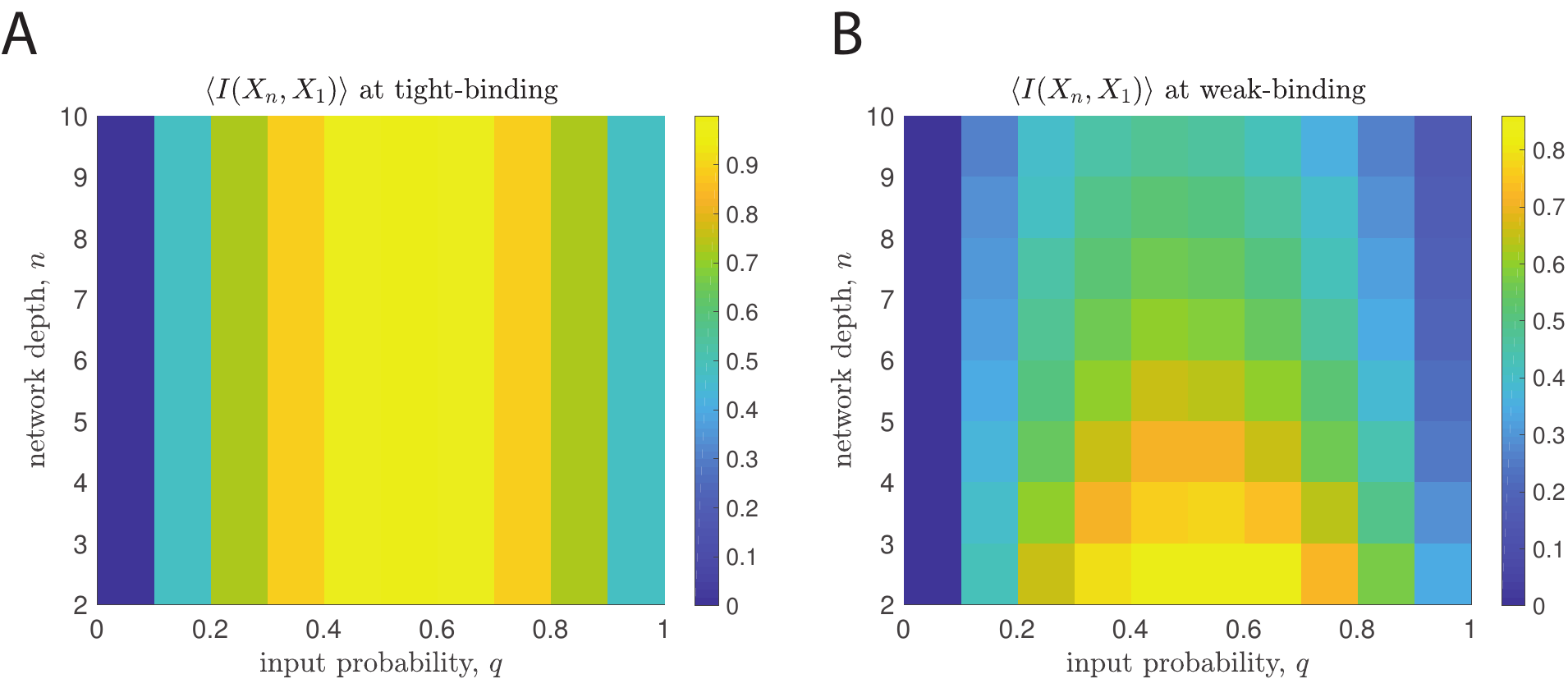}
	\caption{Data processing inequality and biochemical noise: Mutual information across a linear network ($X_1\rightarrow X_2\rightarrow\cdots \rightarrow X_n$) is shown as the color map. Models are described in details in SI Sec.\ref{sec:SImodel}. For tight/weak-binding (left/right panel), mean binding affinity $\beta\mu= -5/-0.1$. In all panels, $q=\P(x_1=1)$ and standard deviation of binding affinities $\sigma = 0.1$.}
	\label{fig:SI-DPI}
\end{center}
\end{figure}

	Here we examine how the depth of network affects information transduction capacity. According to data processing inequality (DPI)\citep{cover2012elements, kinney2014equitability}, information is never gained when transmitted through some noisy channel (or observation process). Formally, DPI states that suppose we have a Markov chain: $X_1\rightarrow X_2\rightarrow X_3$, where $X_1\perp X_3\, \Vert X_2$ (i.e. $X_1$ and $X_3$ are independent conditionally on $X_2$), then it must be that $I(X_1; X_3)\le I(X_1;X_2)$. The pertinent question is therefore how much information degradation across signaling circuit is controlled by biochemical noise due to non-specific PPIs. In Fig.\ref{fig:SI-DPI}, we calculated mutual information for networks described in SI Sec.\ref{sec:SImodel} of varying depth at two binding scenarios. At tight-binding, the noise due to promiscuity of PPIs is small and we observe that DPI is almost saturated (i.e. equality in DPI holds). In the other limit, information is always degraded when as it is relayed downstream.

	\section{Implementation of complex networks}
	\label{sec:SI-2n2}
Here we consider all-to-all connected 2-$n_w$-2 networks, where $n_w\in\mathbb{N}$ is the number of intermediate nodes (see Figure~\ref{fig:infomax}B for an illustration). The goal is to compare the maximum mutual information of circuits with different $n_w$ by using InfoMax (see Algorithm~\ref{euclid}). To do so, we first construct such networks with varying $n_w$, subject them to inputs with different correlations, then maximize mutual information with respect to binding affinities. From a practical point of view, it is useful to define the followings. In the sequel, we use Roman letters $i,j,k,\cdots$ to denote the identity of nodes (i.e. protein species label) while reserving Greek letters $\mu,\nu,\sigma,\cdots$ for configurations (i.e. joint protein phosphorylation states). 

Let $\mbf{x}= (x_1, x_2 )$ be the input (phosphorylation) state vector, $\mbf{y}=(y_1,y_2,\cdots, y_{n_w})$ be the intermediates, and $\mbf{z}=(z_1,z_2)$ be the outputs.  Denote the binding affinity between $y_i$ and $x_j$ as $\theta_{i,j}$ and that between $z_k$ and $y_i$ as $\eta_{k,i}$ (all measured in units of $k_B T$). In other words, these energetic parameters can be summarized by the binding matrix $\bd{\theta}\in\mathbb{R}^{n_w\times 2}$ and $\bd{\eta}\in\mathbb{R}^{2\times n_w}$. To distinguish the variable space (indexed by $i,j,k\cdots$) from the configuration space (indexed by $\mu,\nu,\sigma,\cdots$), let $\mathcal{X} =\{\mbf{x}^{(\mu)}, \mu=1,\cdots, 2^2\}$, $\mathcal{Y} =\{\mbf{y}^{(\mu)}, \mu=1,\cdots, 2^{n_w}\}$, and $\mathcal{Z} =\{\mbf{z}^{(\mu)}, \mu=1,\cdots, 2^2\}$ be the set of input, intermediate, output configurations, respectively. Define $\mathcal{P}_{\mu\nu}: \mathcal{X}\rightarrow\mathcal{Y}$ as a matrix that relates the joint states of two inputs (of dimensionality four) to that of the intermediates and $\mathcal{Q}_{\sigma\mu}: \mathcal{Y}\rightarrow \mathcal{Z}$ for that between the intermediates and the outputs. In matrix form,

\be
\mathcal{P}_{\mu\nu} =\begin{pmatrix}
	1 & \vdots & \vdots & \vdots \\
	0 & \mathcal{P}_{\mu, 2} & \mathcal{P}_{\mu, 2} & \mathcal{P}_{\mu, 2} \\
	\vdots & \vdots & \vdots & \vdots \\
	0& \vdots & \vdots & \vdots
\end{pmatrix},\quad \text{and}\quad	
\mathcal{Q}_{\sigma\mu} =\begin{pmatrix}
	1 & 0 & \hdots & 0\\
	\hdots & \mathcal{Q}_{2,\mu} & \hdots & \hdots \\
	\hdots & \mathcal{Q}_{4,\mu} & \hdots & \hdots \\
	\hdots & \mathcal{Q}_{3,\mu} & \hdots & \hdots 
\end{pmatrix},
\ee
from which one can calculate the input-output and output marginal probability as 
\bea
\mathbb{P}_{\sigma\nu}(\mathcal{Z}|\mathcal{X}) &=& \mathcal{Q}_{\sigma\mu}\mathcal{P}_{\mu\nu}\\
\mathbb{P}_{\sigma}(\mathcal{Z}) &=& \mathcal{Q}_{\sigma\mu}\mathcal{P}_{\mu\nu}q_{\nu},
\eea
where $q_\nu$ is the joint probability of the inputs, namely, $\mathbb{P}(x_1, x_2)$. Note that we use the Einstein notation where repeated indices are implicitly summed over. To simplify notation, denote $y_i^{(\mu)}$ as the $i$-th component of the phosphorylation state vector of intermediate nodes $\mbf{y}^{(\mu)}=(y_1^{(\mu)},\cdots, y_{n_w}^{(\mu)})$. Let $u_\mu=\{i| i\in\mathcal{Y}, y_i^{(\mu)}=1\}$ be the set of intermediate nodes that are phosphorylated and $v_{\mu} = \mathcal{Y}\setminus u_{\mu}$ be those of that are not. The matrix element of $\mathcal{P}_{\mu\nu}$ is therefore:
\bea
\mathcal{P}_{\mu,2} &=&\prod_{i\in u_{\mu}}f(\theta_{i,1})\prod_{j\in v_{\mu}}[ 1- f(\theta_{j,1})]\\
\mathcal{P}_{\mu,3} &=&\prod_{i\in u_{\mu}}f(\theta_{i,2})\prod_{j\in v_{\mu}}[ 1- f(\theta_{j,2})]\\
\mathcal{P}_{\mu,4} &=&\prod_{i\in u_{\mu}}g(\theta_{i,1}, \theta_{i,2})\prod_{j\in v_{\mu}}[ 1- g(\theta_{i,1}, \theta_{i,2})],
\eea
where 
\bea
f(\zeta)&=&\frac{e^{-\zeta}}{1+ e^{-\zeta}}\\
g(\zeta,\xi)&=&\frac{e^{-\zeta}+ e^{-\xi}}{1+ e^{-\zeta}+ e^{-\xi}}.
\eea
Note that in writing down $g$, we ignored higher-order interactions such as those due to cooperativity $\sim e^{-\zeta-\xi}$ etc. We also choose $\mbf{x}^{(\nu)} = \{(0,0), (1,0), (0,1), (1,1)\}$ for $\nu = 1, 2, 3, 4$ ordering. Similarly, 
\bea
\mathcal{Q}_{2,\mu} &=& h(\eta_{1, u_{\mu}}) \, [1- h(\eta_{2, u_{\mu}})]\\
\mathcal{Q}_{3,\mu} &=& [1- h(\eta_{1, u_{\mu}})]\, h(\eta_{2, u_{\mu}}) \\
\mathcal{Q}_{4,\mu} &=& h(\eta_{1, u_{\mu}}) \, h(\eta_{2, u_{\mu}}),
\eea
where the function $h$, with higher order interactions ignored, reads
\be
h(\eta_{i,\mathcal{S}_i})=\frac{\sum_{j\in\mathcal{S}_i}e^{-\eta_{i,j}}}{1+ \sum_{j\in\mathcal{S}_i}e^{-\eta_{i,j}}}
\ee

With all these matrices defined, we can compute the mutual information Eq. \eqref{eq:MI} by a series of matrix multiplications as we did in SI Appendix~\ref{sec:SImodel}.


	\section{Effects of input correlations and pathway cross-talks on information capacity}
	In this appendix, we show the results of a full analysis on pathway cross-talks to complement Figure~\ref{fig:xtalk}. The network studied are depicted as labeled according Figure~\ref{fig:xtalk}A.

	\begin{figure}
		\begin{center}
	\includegraphics[width=0.9 \textwidth]{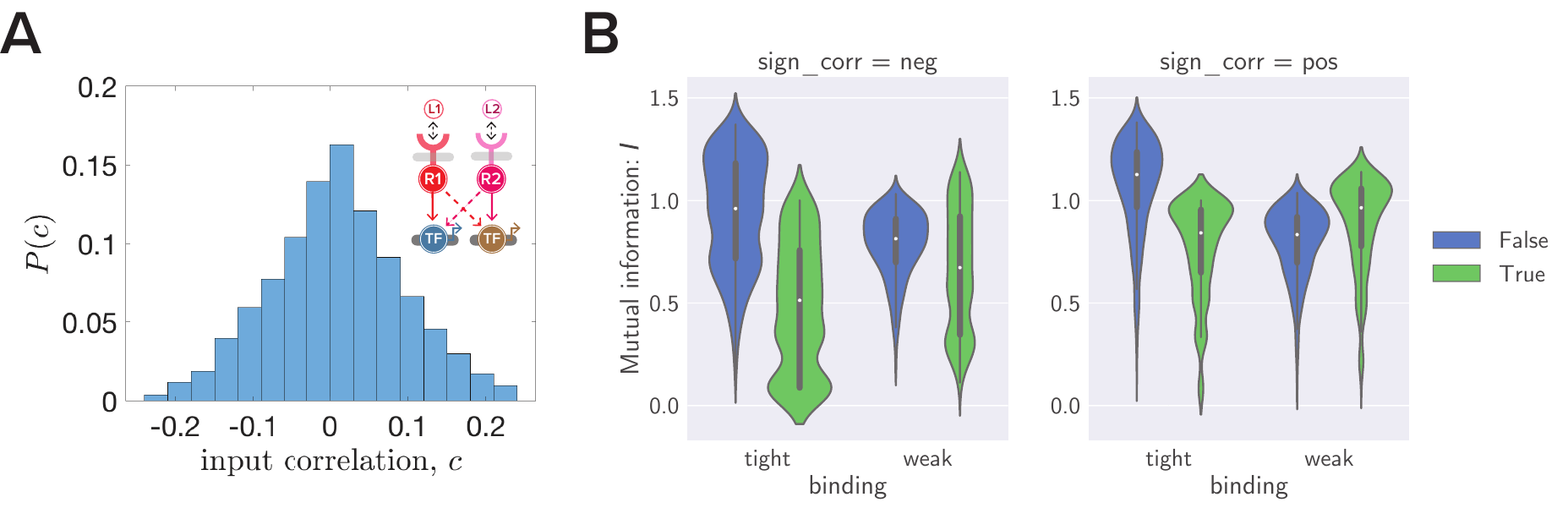}
	\caption{(A) Network studied and the distribution of input correlation $P(c)$ (B) Violin plot for mutual information for network and input correlation shown in A. Violins are categorized according to their cross-talk levels and binding affinities. Blue for 'cross-talk' with $\beta \eta = -5.0$ and green for 'no cross-talk' with $\beta\eta =0$ (see SI Section 4 for details). Tight-binding refers to $\beta\langle\theta\rangle =-5.0$ while weak-binding to  $\beta\langle\theta\rangle =-1.0$.  Note that $P(c)$ is constructed based on Eq.\eqref{eq:corr}. See \emph{Materials and Methods} for details.}
	\label{fig:SI-2by2stat}
\end{center}
\end{figure}

	\begin{figure}
		\begin{center}
	\includegraphics[width=0.3 \textwidth]{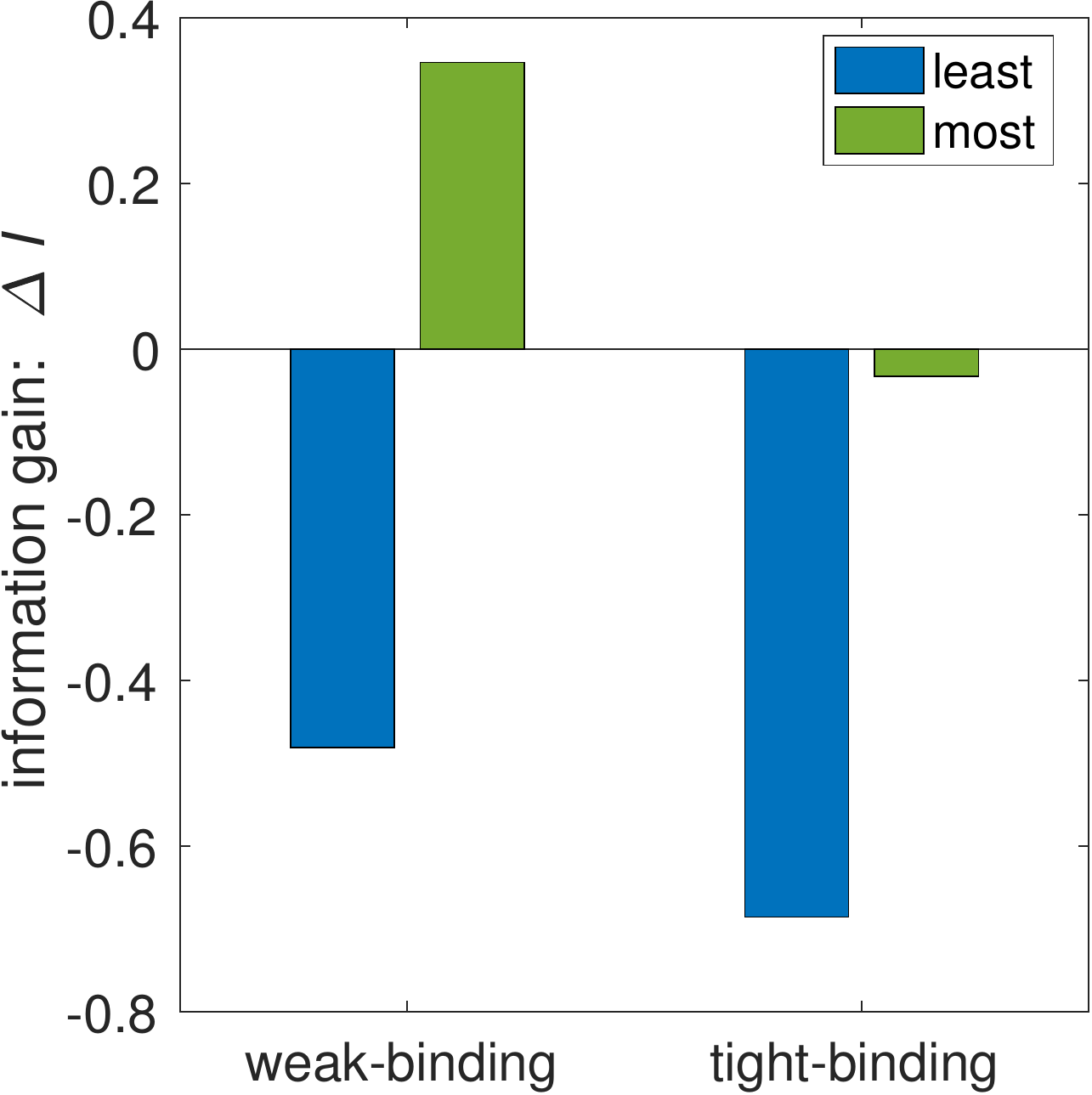}
	\caption{Gain of mutual information through cross-talks for the network shown in Figure~\ref{fig:SI-2by2stat}. Information gain $\Delta I$ is defined as the different between the mutual information with and that without cross-talk. The label 'most' and 'least' are annotated based on maximizing and minimizing $\Delta I$ with respect to input correlations $c$, respectively.  }
	\label{fig:SI-2by2-bw}
\end{center}
\end{figure}

\end{document}